\documentclass[12pt,preprint]{aastex}

\shorttitle{}
\shortauthors{Shinn et al.}

\newif\ifmod
\modfalse
\newif\ifdel
\delfalse
\usepackage{xcolor}
\usepackage{ulem}
\usepackage{censor}
\newcommand{\kms}{km s$^{-1}$}

\newcommand{\Msyr}{$M_\odot$ yr$^{-1}$}

\newcommand{\um}{$\mu$m}

\newcommand{\astH}[1]{H {\small #1}}

\newcommand{\Htwo}{H$_2$}
\newcommand{\HtwolineK}{H$_2$ $\upsilon=1\rightarrow0$ $S$(1)}

\newcommand{\astFe}[1]{Fe {\small #1}}

\newcommand{\Rv}{$R_V$}
\newcommand{\Av}{$A_V$}

\newcommand{\NH}{$N_{\textrm{\tiny{H}}}$}

\newcommand{\Mout}{$\dot{M}_{out}$}

\newcommand{\Macc}{$\dot{M}_{acc}$}

\newcommand{\delref}[1]{\ifdel\textcolor{blue}{#1}\else\fi}
\newcommand{\mod}[1]{\ifmod\textcolor{magenta}{#1}\else{#1}\fi}
\newcommand{\del}[1]{\ifdel\textcolor{blue}{\sout{#1}}\else\fi}
\newcommand{\Moutrg}{{$\sim1\times10^{-6}-4\times10^{-5}$ \Msyr}}
\newcommand{\pkflux}{$\sim3\times10-1\times10^2$ mJy/Beam}

\begin{document}

\title{[\astFe{II}] 1.64 \um{} Imaging Observations of the Outflow Features around Ultracompact \astH{II} Regions in the 1st Galactic Quadrant}

%\author{Jong-Ho Shinn\altaffilmark{1} et al.}
\author{Jong-Ho Shinn\altaffilmark{1}, Kee-Tae Kim\altaffilmark{1}, Jae-Joon Lee\altaffilmark{1,2}, Yong-Hyun Lee\altaffilmark{3}, Hyun-Jeong Kim\altaffilmark{3}, Tae-Soo Pyo\altaffilmark{4}, Bon-Chul Koo\altaffilmark{3}, Jaemann Kyeong\altaffilmark{1}, Narae Hwang\altaffilmark{1}, Byeong-Gon Park\altaffilmark{1,2}}
\email{jhshinn@kasi.re.kr}
\altaffiltext{1}{Korea Astronomy and Space Science Institute, 776 Daeduk-daero, Yuseong-gu, Daejeon, 305-348, Korea}
\altaffiltext{2}{Korea University of Science and Technology, 217 Gajeong-ro, Yuseong-gu, Daejeon, 305-350, Korea}
\altaffiltext{3}{Dept. of Physics and Astronomy, Seoul National University, 1 Gwanak-ro, Gwanak-gu, Seoul, 151-742, Korea}
\altaffiltext{4}{Subaru Telescope, National Astronomical Observatory of Japan, 650 North A`oh\={o}k\={u} Place, Hilo, HI 96720, U.S.A.}

\begin{abstract}
We present \del{the outflow features} \mod{[\astFe{II}] 1.644 \um\ features} around ultracompact H II regions (UCHIIs) found on a quest for the ``footprint'' outflow features of UCHIIs---the features produced by outflowing materials ejected during an earlier, active accretion phase of massive young stellar objects \mod{(MYSOs)}. 
We surveyed 237 UCHIIs in the 1st Galactic quadrant, employing the CORNISH UCHII catalog and UWIFE data, which is an imaging survey in [\astFe{II}] 1.644 \um{} performed with UKIRT-WFCAM under $\sim0.8''$ seeing condition. 
The [\astFe{II}] \del{outflow} features were found around five UCHIIs, one of which has low plausibility. 
We interpret the [\astFe{II}] features to be shock-excited \mod{by outflows from YSOs}, and estimate the outflow mass loss rates from the [Fe II] flux which are \Moutrg.
%The outflow mass loss rates estimated from the [Fe II] flux are \Moutrg. 
We {propose} that the [\astFe{II}] features might be the ``footprint'' outflow features, but more studies are required to clarify it.
{This is} based on the morphological relation between the [\astFe{II}] and 5 GHz radio features, the outflow mass loss rate, the travel time of the [\astFe{II}] features, and the existence of  several YSO candidates near the UCHIIs.
%Although the sample number is small, the UCHIIs accompanying the [\astFe{II}] features have a higher peak flux density, and the outflow mass loss rate show a weak anticorrelation with the peak flux density.
The UCHIIs accompanying the [\astFe{II}] features have \mod{relatively} higher peak flux densities.
\del{The outflow mass loss rate shows no significant correlation with the peak flux density, being limited by the small number of data.}The fraction of UCHIIs accompanying the [\astFe{II}] features, 5/237, is small when compared to the $\sim90$ \% detection rate of high-velocity CO gas around UCHIIs.
We discuss some possible explanations for the low detection rate.
\end{abstract}


\section{Introduction \label{intro}}
The formation of massive stars ($M\ga8\,M_{\sun}$) is still unclear in many aspects \citep{Zinnecker(2007)ARA&A_45_481}.  
One question to be resolved is how massive stars obtain their mass.
It has been increasingly reported that, as in low-mass stars, the disk-mediated
accretion process seems to be underway in forming massive stars \citep[e.g.][]{Beuther(2002)A&A_383_892,Wu(2004)A&A_426_503,SanJose-Garcia(2013)A&A_553_A125,Cooper(2013)MNRAS_430_1125}.
However, it is still uncertain if disk accretion works in the mass range of $M\ga25\,M_{\sun}$ \citep{Zinnecker(2007)ARA&A_45_481}.

One way to grasp the accretion process of massive young stellar objects (MYSOs) is by tracing the outflow features.
In order for the massive (proto)star to accrete mass, the angular momentum of infalling material must be removed.
If not, the angular momentum of infalling material would keep piling on the massive (proto)star, \del{which}\mod{and the (proto)star} would rotate in ever-increasing velocity.  
The outflow plays a significant role in removing this angular momentum \citep{Lada(1985)ARA&A_23_267,Bachiller(1996)ARA&A_34_111}, and produces outflow features in and around the MYSO.
\del{Therefore, we can study the history of accretion process by tracing these ``footprint'' outflow features around ``late-stage'' MYSOs.}\mod{Therefore, by tracing these outflow features we can study the MYSO accretion history. In principle, we can study ``early-stage'' accretion activity even when the MYSO has already evolved to the ``late-stage,'' using the outflow features.}

\del{Ultracompact \ion{H}{2} regions (UCHIIs) are thought to be the late stage of MYSOs, which are small ($\la0.1$ pc), dense ($\ga10^4$ cm$^{-3}$), and no longer accreting significant mass }\delref{\citep{Churchwell(2002)ARA&A_40_27,Zinnecker(2007)ARA&A_45_481}.}\mod{Ultracompact \ion{H}{2} regions (UCHIIs; size $\la0.1$ pc, density $\ga10^4$ cm$^{-3}$) are thought to be the late stage of MYSOs and no longer accreting significant mass \citep{Churchwell(2002)ARA&A_40_27,Zinnecker(2007)ARA&A_45_481}.}
Since UCHII's natal clumps are not completely destroyed yet, one may expect that the materials ejected during the past, active accretion phase are still producing shocked \del{outflow }features around the UCHIIs, {colliding with the natal clump}.
In addition, shocked features can also be produced from the internal working surfaces of jets and outflows \citep{Reipurth(2001)ARA&A_39_403,Arce(2007)2007prpl.conf__245}.
\mod{We will call these shocked features ``footprint'' outflow features, emphasizing the time difference between the current MYSO stage (i.e.~UCHII) and the past MYSO stage when the outflowing material was launched. We thus expect high outflow mass loss rates from the ``footprint'' outflow features, since the outflowing material was launched during the past, active accretion phase.}

\del{These shocked} \mod{These ``footprint'' outflow} features are observable through radiative cooling lines, such as [\astFe{II}] 1.64 \um, H$_2$ 2.12 \um, and CO radio lines \citep{Hollenbach(1989)ApJ_342_306,Neufeld(1989)ApJ_344_251,Kaufman(1996)ApJ_456_611,Wilgenbus(2000)A&A_356_1010,Flower(2010)MNRAS_406_1745}. 
Indeed, CO outflow features have been observed around UCHII regions, such as G5.89-0.39 \citep{Watson(2007)ApJ_657_318,Wood(1989)ApJS_69_831}, G18.67+0.03 \citep{Cyganowski(2012)ApJ_760_L20}, G25.65+1.05 \citep{Shepherd(1996)ApJ_457_267}, and G240.31+0.07 \citep{Shepherd(1996)ApJ_457_267}.
The dynamical timescale of these CO outflow features is $\ga10^4$ yr, which is comparable \del{or greater than }\mod{to} the typical lifetime of UCHIIs \citep[\mod{$\sim4\times10^4$}\del{$\ga5\times10^4$} yr,][]{Wood(1989)ApJS_69_831,Gonzalez-Aviles(2005)ApJ_621_359} and MYSO jet-phase \citep[\mod{$\sim10^4-4\times10^5$}\del{$\ga4\times10^4$} yr,][]{Mottram(2011)ApJ_730_L33,Guzman(2012)ApJ_753_51}.

However, the CO outflow observations \mod{towards UCHIIs} have been performed with low spatial resolutions greater than several arc seconds, limiting detailed study of accretion history, e.g.~outflow morphology.
\mod{Hence, [\astFe{II}] line observations on the ground, whose typical seeing is $\sim1.0''$, can be useful, although its capability for tracing outflows is limited by the depletion of Fe, and the requirements of high density and high shock velocity.}
The \Htwo{} 2.12 \um{} emission line is not useful for tracing outflow features around UCHIIs\mod{,} which emit intense UV, because the emission line is easily excited by far-UV radiation \citep{Hollenbach(1997)ARA&A_35_179}.
\del{We here present the outflow features around UCHIIs seen at higher spatial resolution ($\sim0.8''$) in [\astFe{II}] 1.64 $\mu$m.}\mod{We here present the [\astFe{II}] 1.64 \um\ features around UCHIIs observed at a spatial resolution of FWHM $\sim0.8''$.}
We employed the UKIRT imaging survey of the first Galactic quadrant in [\astFe{II}] 1.64 $\mu$m, i.e.~the UWIFE survey \citep{Lee(2014)arXiv_06_4271}, with the UCHII catalog from the CORNISH survey \citep{Hoare(2012)PASP_124_939,Purcell(2013)ApJS_205_1}.
Among 237 UCHIIs, five UCHIIs were found to have nearby [\astFe{II}] features, one of which had lower plausibility.
\del{We estimate the outflow mass loss rate from [\astFe{II}] fluxes, which ranges \Moutrg.
Based on several observational facts and estimations, we {propose} that the [\astFe{II}] features might be the ``footprint'' outflow features, but more follow-up studies are required to clarify it.}\mod{We estimate the outflow mass loss rate from [\astFe{II}] fluxes, and discuss whether the [\astFe{II}] features can be the ``footprint'' outflow features.}
The relations between the [\astFe{II}] detection rate\del{, the outflow mass loss rate,} and the UCHII parameters from the CORNISH catalog are also discussed.

\section{Observations and Data Reduction \label{obs-red}} 
We used the [\astFe{II}] imaging data from the UWIFE survey \citep{Lee(2014)arXiv_06_4271}.
This survey covers the 1st Galactic quadrant ($7^{\circ}<l<63^{\circ}$, $|b|<1.5^{\circ}$) using the [\astFe{II}] 1.64 \um{} narrow-band filter.
The [\astFe{II}] filter was installed in the Wide-Field Camera \citep[WFCAM,][]{Casali(2007)A&A_467_777} of the United Kingdom Infrared Telescope (UKIRT).
The WFCAM provides four HgCdTe Rockwell Hawaii-II arrays ($2048\times2048$), each of which has a field of view of $13.7'\times13.7'$.
These four arrays are located off-center, forming a square with a $12.9'$ gap.
With this layout, observing at four discrete positions results in a contiguous area covering 0.75 deg$^2$ on the sky, i.e.~a WFCAM tile.

The UWIFE survey were performed through 2012 and 2013, and the observed tiles are shown as gray shaded tiles in Figure \ref{fig-obs}. 
The two UCHIIs (G061.7207+00.8630 and G065.2462+00.3505) uncovered with the UWIFE survey were separately observed by targeting amid the 2013 campaign (14-Sep-2013).
More on the observations is described in \cite{Lee(2014)arXiv_06_4271}.

Data reduction was implemented by the Cambridge Astronomical Survey Unit (CASU).
The reduction process included dark subtraction, flat-fielding, bias subtraction, and sky correction.
The details are described in \cite{Dye(2006)MNRAS_372_1227}.
The astrometric and photometric calibrations \citep{Hodgkin(2009)MNRAS_394_675} were implemented employing the Two Micron All Sky Survey (2MASS) catalog \citep{Skrutskie(2006)AJ_131_1163}.

\section{Analysis and Results \label{ana-res}}
% outflow morphology
% [Fe II] flux
% outflow mass loss rate from [Fe II] flux
% relation between UKIDSS photometry, CORNISH parameters, and [Fe II] flux

\subsection{\del{[\astFe{II}] Detection of Outflow Features}\mod{[\astFe{II}] Feature Detection} around UCHIIs \label{ana-res-det}}
\begin{figure}
\centering
\includegraphics[clip=true,trim=0mm 0mm 0mm 0mm]{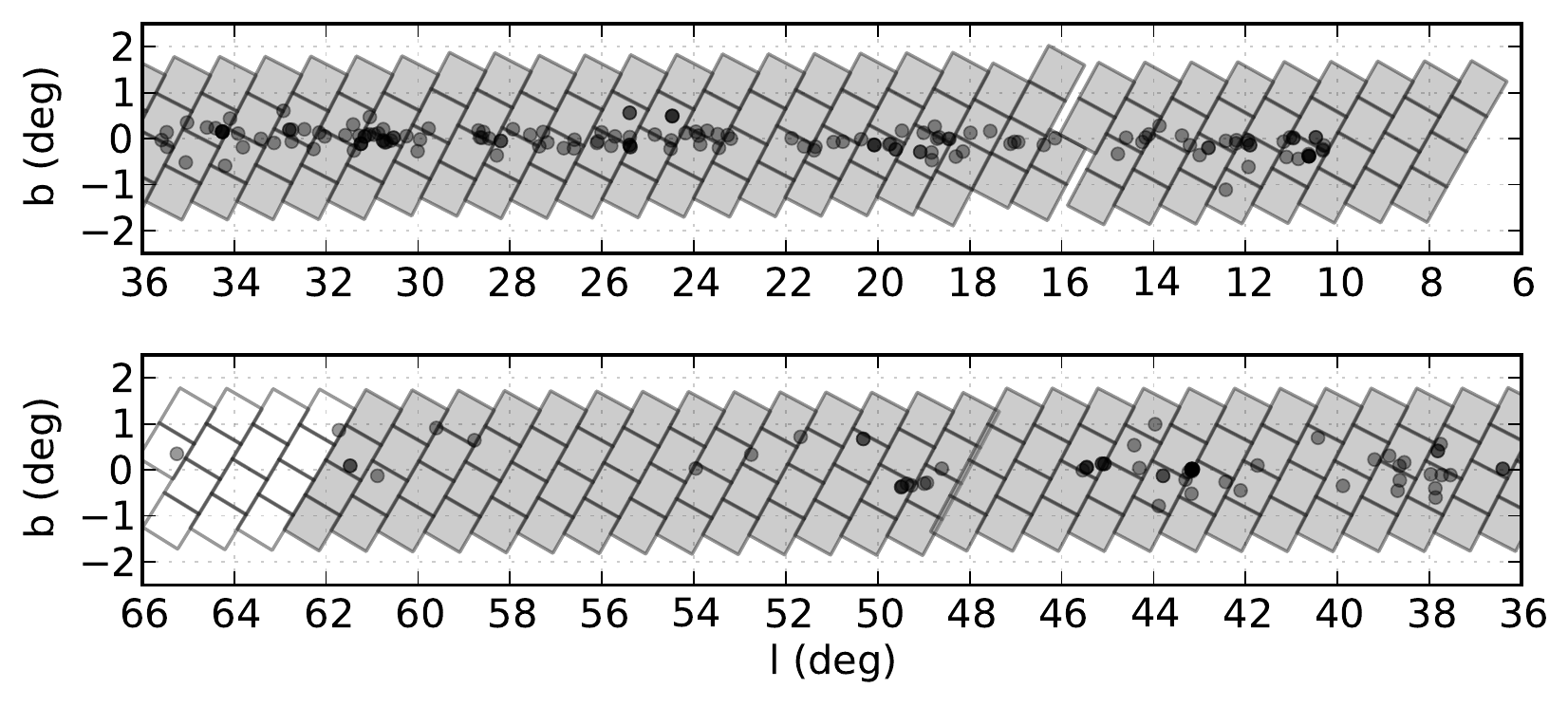}
\caption{
Observed area of the UWIFE survey (\textit{grey squares}) and the position of the UCHII (\textit{grey circles}) in the Galactic Coordinates. 
The position of UCHIIs is from the CORNISH catalog \citep{Purcell(2013)ApJS_205_1}. 
When the positions of circles overlap, the circles appear darker. 
The two UCHIIs (G061.7207+00.8630 and G065.2462+00.3505) uncovered with the UWIFE survey were separately observed by targeting.
\label{fig-obs}
}
\end{figure}

We searched for [\astFe{II}] \del{outflow} features around UCHIIs, using the candidate UCHII catalog  of the CORNISH survey \citep{Hoare(2012)PASP_124_939,Purcell(2013)ApJS_205_1}.
The CORNISH survey was performed with the Very Large Array at 5 GHz over the 1st Galactic quadrant ($10^{\circ}<l<65^{\circ}$, $|b|<1^{\circ}$), which matches well with the coverage of our UWIFE survey (Fig.~\ref{fig-obs}).
The average beam sizes are $\sim1.5''$ and $\sim1.2''$ for the major and minor axes, respectively \citep{Hoare(2012)PASP_124_939}.
The catalog contains 240 sources, and we checked all the sources except three, which had no corresponding H-band image for continuum subtraction.
The unchecked three UCHIIs are G025.3970+00.5614, G025.3983+00.5617, and G025.7157+00.0487.
Figure \ref{fig-obs} shows the distribution of CORNISH UCHII catalog sources plotted over the UWIFE survey coverage.

In order to find [\astFe{II}] \del{outflow} features, we used continuum-subtracted [\astFe{II}] images and RGB composite images of [\astFe{II}] and H (cf.~Fig.~\ref{fig-uchii} and \ref{fig-uchii-cand}).
The H band images are from the UKIDSS Galactic Plane Survey \citep{Lucas(2008)MNRAS_391_136,Lawrence(2007)MNRAS_379_1599}, and the continuum subtraction was done as described in \cite{Lee(2014)arXiv_06_4271}.
We note that there is a time gap of a few years between the [\astFe{II}] and H images.
Therefore, time-variable continuum features can mimic [\astFe{II}] features in the continuum-subtracted and RGB composite images mentioned above.
This is more important considering that some UCHIIs show morphological changes on timescales of years \citep{Acord(1998)ApJ_495_L107,Franco-Hernandez(2004)ApJ_604_L105,vanderTak(2005)A&A_431_993,Galvan-Madrid(2008)ApJ_674_L33}, which suggests changes of the UCHII radiation environment.

We searched a $\sim2.4'\times2.4'$ area around the UCHII sources, and found five UCHIIs that had [\astFe{II}] \del{outflow} features around.
We only picked the [\astFe{II}] features not in contact with the 5 GHz radio features (cf.~Fig.~\ref{fig-uchii}) in order to exclude [\astFe{II}] features produced by the expanding \astH{II} regions.
We also excluded pointlike [\astFe{II}] features, since we cannot tell if they are variable stars or pointlike outflow features, due to the time gap between [\astFe{II}] and H images mentioned above.
These two types of excluded features are shown in Figure \ref{fig-excl} as examples.
Among the picked targets, we filtered out the UCHIIs whose [\astFe{II}] features were less plausible, and named them `candidates'.
One of five selected UCHIIs is classified as a candidate.
Figure \ref{fig-uchii} shows the UCHIIs accompanying [\astFe{II}] \del{outflow} features, while Figure \ref{fig-uchii-cand} shows the candidate.
The CORNISH 5 GHz image with a super-resolution of $\sim1.5''$ \citep{Purcell(2013)ApJS_205_1} is also shown for reference.
More specific descriptions for individual objects are given below.

\subsubsection{G025.3809$-$00.1815 and G025.3824$-$00.1812}
Figure \ref{fig-uchii}a shows the [\astFe{II}] \del{outflow} features around two UCHIIs: G025.3809-00.1815 and G025.3824-00.1812. 
{These UCHIIs are also cataloged in \cite{Thompson(2006)A&A_453_1003}.}
These two UCHIIs reside within the massive star cluster W 42, whose bright central star shows a MK type spectrum of O5--O6 \citep{Blum(2000)AJ_119_1860}.
The [\astFe{II}] features locate southwestward from the two UCHIIs, and show several knotty structures almost in a line pointing towards the two UCHIIs (cf.~white dashed box of Figure \ref{fig-uchii}a). 
We note here the diffuse southeast-northwest continuum features at the upper-left corner of the middle-left panel of Figure \ref{fig-uchii}a, which locate in a symmetric position to the [\astFe{II}] features with respect to the UCHIIs (see also Figure \ref{fig-ccd}a).
We can further trace the [\astFe{II}] features closer to the two UCHIIs in the continuum-subtracted image, although its clear identification is hindered by strong diffuse continuum emissions and the crowded point sources near the two UCHIIs.
The [\astFe{II}] features show no overlap with the radio feature seen in the CORNISH 5 GHz continuum image, hence they seem to have no direct contact.
We note that the 5 GHz morphology of G025.3809$-$00.1815 extends towards the [\astFe{II}] features.

\subsubsection{G028.2879$-$00.3641}
Figure \ref{fig-uchii}b shows the [\astFe{II}] \del{outflow} features around the UCHII G028.2879-00.3641.
{This UCHII is also cataloged in \cite{Kurtz(1994)ApJS_91_659} and \cite{Walsh(1998)MNRAS_301_640}.}
The environment of G028.2879-00.3641 is not crowded as much as G025.3809-00.1815 and G025.3824-00.1812.
The [\astFe{II}] features locate westward from the UCHII, showing an elongated shape (cf.~white dashed box of Figure \ref{fig-uchii}b).
The elongated [\astFe{II}] feature is stretched along the direction roughly perpendicular to the line connecting the UCHIIs and the [\astFe{II}] features.
It is hard to check if any [\astFe{II}] feature exists near the UCHII, since diffuse continuum emissions are so strong.
The [\astFe{II}] features show no overlap with the radio feature seen in the CORNISH 5 GHz continuum image, hence they seem to have no direct contact.
The 5 GHz morphology extends roughly along the direction to the [\astFe{II}] features, as seen in G025.3809$-$00.1815.

\subsubsection{G050.3152+00.6762 and G050.3157+00.6747 \label{ana-res-det-g50}}
Figure \ref{fig-uchii}c shows the [\astFe{II}] \del{outflow} features around two UCHIIs: G050.3152+00.6762 and G050.3157+00.6747.
{These UCHIIs are also cataloged in \cite{Wood(1989)ApJS_69_831}.}
The environment of the two UCHIIs is not crowded as much as G025.3809-00.1815 and G025.3824-00.1812.
The [\astFe{II}] features locate between the two UCHIIs at the northern part, and show knotty or elongated structures (cf.~white dashed box of Figure \ref{fig-uchii}c).
The strong knotty feature seen in the continuum-subtracted image (near the upper-left corner of the white dashed box) also shows a pointlike feature in the H-band image.
This [\astFe{II}] feature could be a fake caused by the temporal variation of the point source as mentioned at the beginning of Section \ref{ana-res-det}.

The elongated [\astFe{II}] feature is stretched along the northeast-southwest direction.
This [\astFe{II}] feature resides at the northwestern edge of the UCHII G050.3157+00.6747 seen in the CORNISH 5 GHz continuum image.
If the elongated [\astFe{II}] feature is related with G050.3157+00.6747, they seem to have a direct physical contact{; in this case, the [\astFe{II}] feature is excluded from our examination, because we only picked [\astFe{II}] features not overlapped with the 5 GHz radio feature (cf.~see above).}
On the other hand, this elongated [\astFe{II}] feature may be related with G050.3152+00.6762, since its stretching orientation is pointing towards G050.3152+00.6762{, and G050.3157+00.6747 shows no [\astFe{II}] feature along its edge except the northwestern edge.}
In that case, the elongated [\astFe{II}] feature can be interpreted as a jetlike feature that has no direct contact with the UCHII.
Unlike previous UCHIIs, the 5 GHz radio feature of G050.3152+00.6762 does not extend towards the [\astFe{II}] features.

\subsubsection{Candidate: G013.8726+00.2818}
Figure \ref{fig-uchii-cand} shows the [\astFe{II}] \del{outflow} features around the UCHII G013.8726+00.2818 (see the arrows).
The [\astFe{II}] features locate at the east, north, and northwest of the UCHII, showing  extended shapes.
The H-band image also shows similar extended features in the area of [\astFe{II}] features.
Therefore, the [\astFe{II}] features could be fakes caused by the temporal variation of the extended continuum emissions; we thus classify this UCHII as a candidate.
The [\astFe{II}] features show no overlap with the crescent radio feature seen in the CORNISH 5 GHz continuum image, hence they seem to have no direct contact.
The crescent shape of the UCHII is well matched with the dark lane seen in the RGB composite image, and extends roughly towards the [\astFe{II}] features.

\subsection{[\astFe{II}] Flux Measurement and Extinction Correction \label{ana-res-flux}}
In order to characterize the physical quantity of the detected [\astFe{II}] features, we measured their fluxes.
The candidate [\astFe{II}] features were excluded from the flux measurements, because they may not be real [\astFe{II}] features.
We picked some regions for measurement, which are shown on the right panels of Figure \ref{fig-uchii}: on-source (\textit{solid line}) and off-source (\textit{dashed line}).
Each region is labeled with a letter of the alphabet in order of increasing RA, and its position is listed in Table \ref{tbl-flux}.
For simplicity of the region name, we just picked one UCHII for the name prefix when there were two UCHIIs.
The measured [\astFe{II}] flux in each region is also listed in Table \ref{tbl-flux}.

We then corrected the extinction effect on the measured [\astFe{II}] flux employing the extinction curve of ``Milky Way, \Rv =3.0'' \citep{Weingartner(2001)ApJ_548_296,Draine(2003)ARA&A_41_241}.
The amount of extinction was estimated by obtaining the total H column density (\NH) from the color excess.
The extinction and the corrected [\astFe{II}] flux are listed in Table \ref{tbl-flux}.
The following sections describe how we obtained the extinction for the individual targets.
The extinctions are $A_V\sim9-20$, which is relatively small compared to the typical value of UCHIIs, $A_V\sim30-50$ \citep{Hanson(2002)ApJS_138_35}.

\subsubsection{G025.3809$-$00.1815 and G025.3824$-$00.1812}
We adopted the color excess $E(H-K)$ estimated from the assumption that the massive stars of the cluster are on the main-sequence \citep{Blum(2000)AJ_119_1860}.
The observed and intrinsic colors are $(H-K)=0.637$ and $(H-K)_0=-0.05$ in the CIT system, respectively.
We used the effective wavelengths for Vega in the CIT system \citep{Bessell(1988)PASP_100_1134} in deriving the total H column density.
$A_{Fe II}$ is about 1.64 mag.

\subsubsection{G028.2879$-$00.3641}
We estimated the color excess $E(H-K)$ using the UKIDSS photometry data \citep{Lucas(2008)MNRAS_391_136}.
We picked the point source which corresponds to the UCHII catalog position, and assumed that the extinction towards this source and the [\astFe{II}] features were the same.
The color of the picked UKIDSS source is $(H-K)=0.556$.
We assumed that this source was on the main-sequence, and adopted the intrinsic color of $(H-K)_0=-0.04$ \citep{Koornneef(1983)A&A_128_84}.
These two colors were compared in the 2MASS system, using the relations in \cite{Lucas(2008)MNRAS_391_136} and \cite{Carpenter(2001)AJ_121_2851}.
We used the isophotal wavelengths of the 2MASS system \citep{Cohen(2003)AJ_126_1090} in deriving the total H column density.
$A_{Fe II}$ is about 1.60 mag.

\subsubsection{G050.3152+00.6762 and G050.3157+00.6747}
We estimated the color excess $E(J-H)$ using the UKIDSS photometry data \citep{Lucas(2008)MNRAS_391_136}.
We picked two point sources which respectively correspond to the two UCHII catalog positions.
The colors of the picked UKIDSS sources are $(J-H)=1.196$ for G050.3157+00.6747 and $(J-H)=2.054$ for G050.3152+00.6762, respectively.
We used the mean value for the observed color.
The color $(J-H)$ is used rather than $(H-K)$, because the UKIDSS photometry data do not provide the K-band photometry.
We assumed that the two sources were on the main-sequence, and adopted the intrinsic color of $(J-H)_0=-0.13$ \citep{Koornneef(1983)A&A_128_84}.
These two colors were compared in the 2MASS system, using the relations in \cite{Lucas(2008)MNRAS_391_136} and \cite{Carpenter(2001)AJ_121_2851}.
We used the isophotal wavelengths of the 2MASS system \citep{Cohen(2003)AJ_126_1090} in deriving the total H column density.
$A_{Fe II}$ is about 3.71 mag.

{\subsection{Excitation Mechanism for the [\astFe{II}] Features \label{ana-res-mec}}}
%\subsubsection{G025.3809$-$00.1815 and G025.3824$-$00.1812}
%\subsubsection{G028.2879$-$00.3641}
%\subsubsection{G050.3152+00.6762 and G050.3157+00.6747}
In this section, we examine the excitation mechanism for the observed [\astFe{II}] features.
We think the [\astFe{II}] features are probably excited by shocks rather than UV radiation\mod{, as listed below}.
\mod{The shock driver is likely to be outflows from YSOs, because, in star forming regions, they are the most probable source generating supersonic motions that result in the observed [\astFe{II}] features.}

First, the [\astFe{II}] features show different morphology from the diffuse near-infrared continuum features, which indicate the irradiated area\footnote{This near-infrared continuum is almost certainly dust-scattered light. It is not likely to be from thermal dust, since the temperature should be as high as $\sim10^3$ K. This high dust temperature is not easy to achieve in the photo-dissociation region \citep[cf.][]{Hollenbach(1997)ARA&A_35_179}. Such a temperature is observed at supernovae \citep[e.g.][]{Fox(2009)ApJ_691_650}.}.
The dominant UV radiation sources in the region are the massive stars ionizing the UCHIIs, and it is known that the [\astFe{II}] emission well traces the warm neutral zone in the photo-dissociation regions \citep{Burton(1990)ApJ_365_620}.
If the [\astFe{II}] features are radiatively excited, they should show features similar to the continuum features.
However, this is not the case (cf.~Figures \ref{fig-uchii} and \ref{fig-uchii-ukidss}).
Additionally, in the cases of G025.3809$-$00.1815, G025.3824$-$00.1812 (Figures \ref{fig-uchii}a and \ref{fig-uchii-ukidss}a), and G028.2879$-$00.3641 (Figures \ref{fig-uchii}b and \ref{fig-uchii-ukidss}b), there are continuum features that locate closer to the UCHIIs than the observed [\astFe{II}] feature.
These continuum features have no corresponding [\astFe{II}] feature, which means the radiatively excited [\astFe{II}] is weak.
Therefore, the observed [\astFe{II}] features have even less probability of being radiatively excited.
We note that G050.3157+00.6747 (the left UCHII in Figure \ref{fig-uchii}c and \ref{fig-uchii-ukidss}c) is excluded from this examination, because we only considered the link between G050.3152+00.6762 (the right UCHII in Figure \ref{fig-uchii}c and \ref{fig-uchii-ukidss}c) and the [\astFe{II}] feature (cf.~section \ref{ana-res-det-g50}).

Second, the observed [\astFe{II}] features locate far from the ionized regions, i.e.~UCHIIs, as can be seen from the 5 GHz radio images (Figure \ref{fig-uchii}).
Therefore, the intensity of the hydrogen recombination line at the location of observed [\astFe{II}] features would be low.
Considering that the photo-ionized gas shows the line ratio of [\astFe{II}]/Br$\gamma\simeq0.1-2.5$ \citep{Alonso-Herrero(1997)ApJ_482_747}, the intensity of the photoionzed [\astFe{II}] line would also be low at the location of the observed [\astFe{II}] features.

\subsection{Estimation of Outflow Mass Loss Rate \label{ana-res-mout}}
Based on the argument that the [\astFe{II}] features are probably shock-excited \mod{by outflows from YSOs} (cf.~Section \ref{ana-res-mec}), we estimate the outflow mass loss rate (\Mout) from the [\astFe{II}] flux {(Table \ref{tbl-flux})} in two different ways \citep[cf.~Section 3.2 of][]{Shinn(2013)ApJ_777_45}.
We call them the ``Fe-Shell'' and ``Fe-Stream'' methods, respectively.

\del{The ``Fe-Shell'' method assumes that the [\astFe{II}] feature is excited at the shocked shell of wind and ambient material.}\mod{The ``Fe-Shell'' method assumes that the [\astFe{II}] feature is excited by the wind shock and the ambient shock, both of which are J-type \citep[for shock types, see][]{Draine(1993)ARA&A_31_373}, when the outflowing material is colliding with the ambient medium (cf.~Fig.~3 of \citealt{Shinn(2013)ApJ_777_45}). We estimate \Mout{} through the shock luminosity. The shock luminosity, which is expressed in terms of \Mout, is derived from the [\astFe{II}] flux employing the shock model calculation.}
The ``Fe-Stream'' method assumes that the [\astFe{II}] feature is a well-collimated stream of \del{excited}\mod{ionized} gas flowing from the outflow source.
\mod{We derive the total mass of the collimated medium from the [\astFe{II}] flux, assuming the typical excitation condition of [\astFe{II}] line. Then, we estimate the travel time from the measured outflow length and the assumed outflow velocity. From these mass and time values, we estimate \Mout.}

We applied the ``Fe-Shell'' and ``Fe-Stream'' methods to the knotty and longish [\astFe{II}] features, respectively.
The ``Fe-Stream'' method is only applied to the [\astFe{II}] feature in Figure \ref{fig-uchii}c, relating it with the UCHII G050.3152+00.6762.
\mod{The Fe depletion onto dust grains is also considered for both methods, in order to reflect the Fe depletion in jets \citep[e.g.][]{Beck-Winchatz(1996)AJ_111_346,Mouri(2000)ApJ_534_L63,Nisini(2002)A&A_393_1035,Nisini(2005)A&A_441_159,Podio(2006)A&A_456_189,Giannini(2008)A&A_481_123,Giannini(2013)ApJ_778_71,Antoniucci(2014)A&A_566_A129}. The logarithmic abundance of $-4.63$ \citep{Allen(2008)ApJS_178_20}, which is depleted by $0.13$ dex \citep{Asplund(2009)ARA&A_47_481}, is used for the ``Fe-Shell'' method. We apply the same depleted abundance for the ``Fe-Stream'' method, i.e.~$A_{Fe/H}=2.3\times10^{-5}$ (cf.~eq (10) of \citealt{Shinn(2013)ApJ_777_45}).}
Table \ref{tbl-out} lists the method used and the results, and \Mout{} ranges \Moutrg.

In applying the ``Fe-Shell'' method to G025.3809$-$00.1815-B and G025.3809$-$00.1815-C (Figure \ref{fig-uchii}a), we modify Eq.~(8) of \cite{Shinn(2013)ApJ_777_45}.
\del{These features have the corresponding \Htwo{} features (Figure \ref{fig-uchii-ukidss}a), and hence the [\astFe{II}] feature probably comes from the wind-shock only, because C-type (slower) shock usually shows strong \HtwolineK{} 2.12 \um{} emissions }\delref{\citep[cf.][]{Wilgenbus(2000)A&A_356_1010}}\del{ and the ambient shock is likely C-type (slower).}\mod{These [\astFe{II}] features have the corresponding \HtwolineK{} 2.12 \um{} features (Figure \ref{fig-uchii-ukidss}a), and the [\astFe{II}] features probably come from the wind-shock only, rather than both wind and ambient shocks. We think the \Htwo{} feature is likely from C-type shocks rather than J-type shocks, because only some of the observed [\astFe{II}] features accompany the \Htwo{} features. In this sense, the ambient shock is more likely to be C-type than the wind shock, since the ambient shock is propagating into a denser medium.}
\del{Therefore, we use the modified equation $L_{mech}=4/27\,\dot{M}_{out}\,v_w^2$ instead of the eq.~(8) of }\delref{\cite{Shinn(2013)ApJ_777_45}}\del{ in the estimation of \Mout .}\mod{Therefore, we modified Eq.~(7) of \cite{Shinn(2013)ApJ_777_45} to be $L_{mech}=1/2\,\dot{M}_{sw}\,v_{sw}^2$, and hence Eq.~(8) of \cite{Shinn(2013)ApJ_777_45} is modified to $L_{mech}=4/27\,\dot{M}_{out}\,v_w^2$.}
G028.2879$-$00.3641 has some filamentary \Htwo{} features around it (Figure \ref{fig-uchii-ukidss}b), but its morphology is different from that of [\astFe{II}] features (Figure \ref{fig-uchii}b).
We think this \Htwo{} feature is probably excited by radiation, because it closely follows the near-infrared continuum features (cf.~Section \ref{ana-res-mec}).

Table \ref{tbl-out} also lists other physical parameters required for the \Mout{} estimation.
We adopted the outflow velocity of {200} \kms, because it falls within the typical outflow velocity of YSOs \citep{Reipurth(2001)ARA&A_39_403} and the J-shock develops for a shock velocity of $>50$ \kms{} under the typical cloud environments \citep{Draine(1993)ARA&A_31_373,LeBourlot(2002)MNRAS_332_985}.
We adopted 100 \kms{} for G025.3809$-$00.1815-B and G025.3809$-$00.1815-C (Figure \ref{fig-uchii}a and \ref{fig-uchii-ukidss}a), in order to make the ambient shock velocity slow enough to be C-type.
The length scale and solid angle of the [\astFe{II}] feature are from the ellipse used for the [\astFe{II}] flux measurement (cf.~Section \ref{ana-res-flux}).
The distance to the individual UCHIIs was adopted as described in the following sections.
We note that the \Mout{} estimation includes the uncertainties that originate from several assumptions, such as \mod{the Fe depletion,} the outflow velocity and the inclination \citep[cf.~Section 3.2.3 of][]{Shinn(2013)ApJ_777_45}.

\subsubsection{G025.3809$-$00.1815 and G025.3824$-$00.1812}
% 4.3 kpc \citep{Lester(1985)ApJ_296_565}, <-- weak evidence
% 11.5 kpc \citep{Russeil(2003)A&A_397_133}, <-- mis-identification  
% 10.8 kpc \citep{Churchwell(1990)A&AS_83_119}, <-- not a definite estimation
The kinematic distances to the cluster W 42 where the two UCHIIs reside were estimated as follows: 3.8 kpc \citep{Anderson(2009)ApJ_690_706}, 3.92 kpc \citep{Jones(2012)ApJ_753_62}, 4.0 kpc \citep{Kolpak(2003)ApJ_582_756}, 5 kpc \citep{Radhakrishnan(1972)ApJS_24_49}, 10.8 kpc \citep{Churchwell(1990)A&AS_83_119}, 13.4 kpc \citep{Wilson(1972)A&A_19_354}, 13.5 kpc \citep{Downes(1980)A&AS_40_379}.
The spectrophotometric distances were estimated to be 2.2 kpc \citep{Blum(2000)AJ_119_1860} and 2.67 kpc \citep{Moises(2011)MNRAS_411_705}.
The spectrophotometric distance suggests the UCHIIs probably locate at the near side among two ambiguous kinematic distances.
Therefore, we adopted the distance of 3.9 kpc, which is the average of near-side distances estimated within ten years.

\subsubsection{G028.2879$-$00.3641}
% 3.3 kpc \citep{Giveon(2007)AJ_133_639} <-- wrong citation, R_gal rather than R_kin.
The kinematic distances were estimated to be 3.0 kpc \citep{Churchwell(2010)A&A_513_A9,Anderson(2009)ApJ_690_706} and 3.29 kpc \citep{Cyganowski(2009)ApJ_702_1615}.
We adopted 3.2 kpc, averaging these values.

\subsubsection{G050.3152+00.6762 and G050.3157+00.6747}
% 2.1 kpc \citep{Ginsburg(2011)ApJ_736_149} <-- contradictory distances within the paper
% 8.7 kpc \citep{Churchwell(1990)A&AS_83_119}, <-- based on a trend
The kinematic distances were estimated to be 2.1 kpc \citep{Watson(2003)ApJ_587_714}, 2.16 kpc \citep{Araya(2002)ApJS_138_63}, 8.7 kpc \citep{Churchwell(1990)A&AS_83_119}, and 9.7 kpc \citep{Anderson(2009)ApJ_690_706}.
The extinction to these UCHIIs are about \Av{} $\sim$ 20, more than ten magnitude higher than other UCHIIs that locate at around $3-4$ kpc (Table \ref{tbl-flux}).
We think the distance of $\sim2.1$ kpc is unlikely to give such a high extinction.
Therefore, we adopted 9.2 kpc, averaging the two larger estimations.

\subsection{\del{[\astFe{II}] Outflow Features around UCHIIs}\mod{UCHIIs with nearby [\astFe{II}] Outflow Features} and the CORNISH Catalog Parameters}
In this section, we investigate the relations between the [\astFe{II}] outflow features and the CORNISH catalog parameters (Table \ref{tbl-cor}).
First, we contrast the location of the UCHIIs accompanying [\astFe{II}] features in the plots of three UCHII parameters: angular scale, integrated flux density, and peak flux density.
These three UCHII parameters are from the CORNISH catalog \citep{Purcell(2013)ApJS_205_1}, and Figure \ref{fig-cor} shows their scatter plots.
The plot of angular scale versus integrated flux density presents a rough correlation (\textit{black point}).
This correlation seems to reflect the distance effect on the UCHIIs of semi-uniform luminosity and physical size, which should show (angular scale) $\sim$ (distance)$^{-1}$ and (flux) $\sim$ (distance)$^{-2}$, and hence (angular scale) $\sim$ (flux)$^{1/2}$.
The plot of angular scale versus peak flux density presents almost no correlation.
This seems to be caused by the weakening of the distance effect, replacing the integrated flux density with the peak flux density, which is likely independent of the distance.

The UCHIIs accompanying [\astFe{II}] features occupy separate regions in Figure \ref{fig-cor}.
This is more easily seen in the plot between peak flux density and angular size.
The UCHIIs reside over the range of \mod{relatively} higher peak flux density (\pkflux), regardless of angular scale.

\del{In order to investigate if there is any relation between the peak flux density and the outflow mass loss rate, we make a scatter plot (Figure \ref{fig-cor-mout}).
The [\astFe{II}] feature seen in Figure \ref{fig-uchii}a is linked to both UCHIIs (G025.3809$-$00.1815 and G025.3824$-$00.1812), since it is hard to pin down which UCHII is more likely responsible for the [\astFe{II}] features.
Therefore, there are some data points repeated at different y-axis values in Figure \ref{fig-cor-mout}.
Figure \ref{fig-cor-mout} shows a Pearson correlation coefficient of {$\sim-0.50$}, but its significance is low, because the number of data points is small and it is uncertain for some [\astFe{II}] features which UCHII is related with. }

\subsection{Point Sources with Near-infrared Excess around UCHIIs}
We are looking for the ``footprint'' outflow features around UCHIIs which were produced by the material ejected during the past, active accretion phase of MYSOs.
However, considering that massive stars usually form in clusters\mod{,} which include numerous other YSOs \citep[][and references therein]{Zinnecker(2007)ARA&A_45_481}, the outflow features can also be produced by \del{the }other YSOs.
In order to examine the population of YSOs near the UCHIIs, we investigated the near-infrared color excess of nearby point sources.

Figure \ref{fig-ccd} shows the color-color diagram of ($J-H$) versus ($H-K$) and the corresponding composite image of $H$ and [\astFe{II}].
We use the photometry data from the UKIRT Infrared Deep Sky Survey \citep[UKIDSS,][]{Lawrence(2007)MNRAS_379_1599} Galactic Plane Survey \citep[GPS,][]{Lucas(2008)MNRAS_391_136}, except for G028.2879$-$00.3641.
The UKIDSS GPS do not provide the $J$ band photometry for G028.2879$-$00.3641, hence we instead used the photometry data from 2MASS \citep{Skrutskie(2006)AJ_131_1163}, which has inferior spatial resolution and survey depth to UKIDSS GPS \citep{Lawrence(2007)MNRAS_379_1599}.
In the color-color diagram, we plot the locus of main-sequence stars employing the values in \cite{Hewett(2006)MNRAS_367_454}.
For the 2MASS data, we transformed the locus using the equation in \cite{Hewett(2006)MNRAS_367_454}.
The extinction line was calculated using the results of \cite{Rieke(1985)ApJ_288_618}.

We classify the point sources that reside below the extinction line as showing the near-infrared excess.
As Figure \ref{fig-ccd} shows, there are several point sources with near-infrared excess around UCHIIs.
These sources could be YSOs, and they could potentially produce the [\astFe{II}] outflow features observed around UCHIIs.

\section{Discussion \label{dis}}
\subsection{Nature and Origin of the [\astFe{II}] Features around UCHIIs \label{dis-nat}}
% Why yes! Why not!
% morphology radio-feii
% CO observation results of the [FeII]-detected UCHIIs
% outflow mass loss rate of UCHII \cite{Roth(2013)arXiv_11_5912}
% outflow timescale: UCHII-FeII distance, outflow velocity
% nearby YSO origin rather than UCHII

We seek the ``footprint'' outflow features around UCHIIs, which are produced by the material ejected during the prior, active accretion phase of MYSOs.
The [\astFe{II}] 1.644 \um{} emission line was employed, and we found that five out of 237 UCHIIs have nearby [\astFe{II}] features.
Based on the given observational facts and estimations, it seems that the detected [\astFe{II}] features might be the ``footprint'' outflow features, but more detailed and targeted study is required to clarify it.
More specific points are given below.

First, the morphological relation between the [\astFe{II}] and 5 GHz radio continuum features \del{supports}\mod{is compatible with} the ``footprint'' interpretation.
The radio features are elongated towards the [\astFe{II}] features in the cases of G025.3809$-$00.1815 (Figure \ref{fig-uchii}a) and G028.2879$-$00.3641 (Figure \ref{fig-uchii}b).
The candidate G013.8726+00.2818 (Figure \ref{fig-uchii-cand}) also shows a similar morphological relation.
Only G050.3152+00.6762 (Figure \ref{fig-uchii}c) does not show such a relation.
However, considering that the [\astFe{II}] features near G050.3152+00.6762 could be produced by the other UCHII G050.3157+00.6747, all the facts suggest a relation between the [\astFe{II}] and radio continuum features.

This morphological relation can be understood in the disk-accreting MYSO model \citep[e.g.][]{Yorke(2002)ApJ_569_846,Krumholz(2009)Science_323_754,Kuiper(2013)ApJ_763_104}, which has a bipolar cavity excavated by the radiation pressure and outflow materials.
The cavity structure is seen up to the radial distance of $\sim2\times10^4$ AU $\sim0.1$ pc, which is comparable to the typical size of a UCHII \citep{Wood(1989)ApJS_69_831,Hoare(2007)2007prpl.conf__181}.
The ionized gas would expand more rapidly through the cavity, hence the morphology of ionized gas would be extended along the cavity direction, i.e. the outflow direction.
The diffuse near-infrared continuum features seen northeastward from G025.3809$-$00.1815 and G025.3824$-$00.1812 (Figure \ref{fig-uchii}a) likely emanated through such a cavity.
Also, the high-velocity CO gas \del{observed toward G25.65+1.05 and G240.31+0.07 }\delref{\citep{Shepherd(1996)ApJ_457_267}}\mod{mapped in the study of \cite{Shepherd(1996)ApJ_457_267}} are probably from the bipolar outflows \del{running through}\mod{following} the cavity.

Second, \Mout{} does not give a clear answer as to whether the [\astFe{II}] features are the ``footprint'' outflow features or not.
\Mout{} is estimated to be \Moutrg{} (Table \ref{tbl-out}).
If we assume a typical ratio between the outflow mass loss rate and the disk accretion rate \Mout/\Macc=0.1 \citep{Ellerbroek(2013)A&A_551_A5,Ray(2007)2007prpl.conf__231,Frank(2014)arXiv_02_3553}, we can guess \Mout{} from \Macc{} of MYSO models.
The \Macc{} of MYSO models ranges $\sim10^{-4}-10^{-3}$ \Msyr{} \citep[e.g.][]{Yorke(2002)ApJ_569_846,Krumholz(2009)Science_323_754,Krumholz(2012)ApJ_754_71,Kuiper(2013)ApJ_772_61}, hence \Mout{} would be around $\sim10^{-5}-10^{-4}$ \Msyr{}.
Our \Mout{} estimation is overlapped with this model-inferred \Mout{}, but a little lower.
Meanwhile, our \Mout{} is comparable to the \Mout{} of \del{low-mass stars}\mod{low- or intermediate-mass YSOs} such as the Herbig Ae/Be star and FU Ori object \citep{Ellerbroek(2013)A&A_551_A5}.
We therefore\del{, only based on \Mout,} cannot exclude the possibility that the observed [\astFe{II}] features were produced by \del{low-mass stars rather than the massive stars ionizing the UCHIIs}\mod{nearby low- or intermediate-mass YSOs}.

Third, the travel time of the [\astFe{II}] features does not exclude the ``footprint'' interpretation, although the time is roughly estimated.
If \del{we assume that }the outflow material that excites the [\astFe{II}] features is ejected from the \mod{location of} a UCHII, we can guess the time the material spent to arrive at the [\astFe{II}] feature position.
We adopted the velocity and the distance listed in Table \ref{tbl-out}, and assumed the outflow  material flowing on the plane of the sky.
The travel time is estimated to be {$\sim(1-8)\times10^{3}$} yr (Table \ref{tbl-time}).
These times are shorter than the typical lifetime of UCHIIs \citep[$\ga5\times10^4$ yr,][]{Wood(1989)ApJS_69_831,Gonzalez-Aviles(2005)ApJ_621_359} and MYSO jet-phase \citep[$\ga4\times10^4$ yr,][]{Mottram(2011)ApJ_730_L33,Guzman(2012)ApJ_753_51}.
Therefore, depending on when the outflow was launched, the [\astFe{II}] feature can be interpreted as the ``footprint'', or even as \mod{having been} excited by the outflow launched \textit{during} the UCHII phase.
We note that some studies assert there is the ongoing accretion onto the central object even after the \astH{II} region development \citep{Keto(2002)ApJ_568_754,Keto(2003)ApJ_599_1196,Keto(2006)ApJ_637_850,Keto(2008)ApJ_678_L109}.
For reference, we calculated the expansion time of UCHIIs (Table \ref{tbl-time}), using the typical sound velocity of the \astH{II} region and the deconvolved size of UCHIIs (Table \ref{tbl-cor}).
\mod{Both the expansion and travel times have a similar order of magnitude.}
The adopted sound velocity is $c_s=\sqrt{\gamma k T/\mu m_H}\sim15$ \kms{} with $\gamma=5/3$, $\mu=0.61$, and $T=10^4$ K.
We note that the travel time can be longer if the inclination to the plane of sky is allowed for.
The outflow velocity is definitely another element that can make the time shorter or longer.

Fourth, several YSO candidates existing near the UCHIIs hinder the ``footprint'' interpretation.
As Figure \ref{fig-ccd} shows, there are several point sources that have near-infrared color excesses near the UCHIIs.
If these sources are YSOs, \del{then }they could have produced the observed [\astFe{II}] features.
For example, the YSO candidates close to the cataloged UCHII position could have (Figure \ref{fig-ccd}a and b), and the YSO candidate at the northeast of G050.3152+00.6762 also could have (Figure \ref{fig-ccd}c).
The three points argued above---morphology, \Mout, travel time---can be understood without the ``footprint'' interpretation, if the YSO candidates \mod{near UCHIIs} can produce \Mout{} of \Moutrg, like the Herbig Ae/Be star or FU Ori objects \citep{Ellerbroek(2013)A&A_551_A5}.
\mod{In that case, }the travel time \mod{of $\sim(1-8)\times10^3$ yr (Table \ref{tbl-time})} is \del{consistent with}\mod{not contradictory to} the \mod{typical} lifetime of a Herbig Ae/Be star \cite[a few Myr,][]{Waters(1998)ARA&A_36_233} or a FU Ori object \citep[$\sim0.1$ Myr,][]{Hartmann(1996)ARA&A_34_207}.

\subsection{[\astFe{II}] Feature Detection\del{, Outflow Mass Loss Rate,} and UCHII Parameter}
% relation between detectability, outflow rate and peak flux density
% low detectability
% UCHII modeling prediction for peak flux density
% feii feature difference - more statistics - weak feii features
Figure \ref{fig-cor} shows that the UCHIIs accompanying [\astFe{II}] features reside over the \mod{relatively} higher range of peak flux density \mod{(\pkflux)}, regardless of the angular size.
Considering that the peak flux density would decrease as the UCHII expands, this distribution indicates that the [\astFe{II}] features are more likely produced around younger UCHIIs.
This is reasonable, because younger UCHIIs would disperse less nearby natal clump materials and would increase the chance of collision between the outflow and natal clump materials.
\del{Figure \ref{fig-cor-mout} shows no significant correlation between \Mout{} and peak flux density, being limited by the small number of data points.
We could investigate more on this relation by upsizing the statistics.}

{Five out of 237 UCHIIs showed nearby [\astFe{II}] features.
For comparison, we note the $\sim90$\% detection rate of high-velocity CO gas around 94 UCHIIs observed in radio \citep{Shepherd(1996)ApJ_457_267}.
Also, \cite{Varricatt(2010)MNRAS_404_661} detected \Htwo{} features around four \del{UCHIIs out }of 13 UCHIIs, imaging \HtwolineK{} 2.12 \um{} emission line.
The \HtwolineK{} line is well excited by far-UV radiation \citep{Hollenbach(1997)ARA&A_35_179}, hence we should be careful in interpreting these \Htwo{} features as outflow features.
If these \Htwo{} features are outflow features, \del{then }the overall CO, \Htwo, [\astFe{II}]  detection rates can be understood with the probability distribution of outflow velocity.
In order to be excited, CO, \Htwo, and [\astFe{II}] require shock velocities of increasing magnitude.
Therefore, if the outflow velocity distribution has a higher probability at a lower velocity, then the CO, \Htwo{}, [\astFe{II}] detection rates of decreasing order can be explained.   
}

One possible reason for the low [\astFe{II}] detection rate is the extinction towards UCHIIs, which is typically $A_V\ga30-50$ \citep{Hanson(2002)ApJS_138_35} $\sim A_H\ga6-9$ \citep{Draine(2003)ARA&A_41_241}.
Therefore, we may miss numerous [\astFe{II}] features.
Note that $A_V$ for three UCHIIs in Table \ref{tbl-flux} are relatively low ($A_V\sim9-20$).
Another possible reason is that the outflow cavity is already open to the outside of the natal clump \citep[cf.][]{Kim(2001)ApJ_549_979}, and hence less materials exist to collide with the outflow.
The typical size of clumps \citep[$\sim0.3-3$ pc,][]{Bergin(2007)ARA&A_45_339} is {smaller than} the traveling distance of a {$100-200$} \kms{} outflow material over the typical lifetime of a MYSO jet-phase \citep[$\ga4\times10^4$ yr,][]{Mottram(2011)ApJ_730_L33,Guzman(2012)ApJ_753_51}.
If the outflow direction is steady, it would make the situation worse.
Another possible reason is the weakness of [\astFe{II}] features.
The shocked [\astFe{II}] feature are likely to be unresolved and the [\astFe{II}] flux is proportional to the column density along the line of sight.
Hence, weak [\astFe{II}] features from low column density might not have been detected due to the detection limit.
The UWIFE survey has a nominal 5-$\sigma$ detection limit of $\sim18.7$ mag for point sources \citep{Lee(2014)arXiv_06_4271}.
{The low outflow velocity is another possible reason.
If the outflow material is ejected with velocity too low to produce J-type shock \citep[cf.][]{Draine(1993)ARA&A_31_373,LeBourlot(2002)MNRAS_332_985} during a certain amount of time, e.g.~low accretion duration, the chance of seeing the [\astFe{II}] feature is reduced.}
Finally, if the outflow material is clumpy or bullet-like rather than continuous stream-like, then the chance to observe a shocked [\astFe{II}] feature would be lower, depending on the emergence frequency of the outflow material.
The observed [\astFe{II}] features appear to be both stream-like (Figure \ref{fig-uchii}c) and clumpy (Figure \ref{fig-uchii}b) outflow materials.
More observations for the weaker [\astFe{II}] features would help in assessing the effects of outflow type on the [\astFe{II}] feature detection.

\section{Conclusions \label{concl}}
We sought the ``footprint'' outflow features around UCHIIs which are produced by materials ejected during the past, active accretion phase of MYSOs.
The UWIFE survey data \citep{Lee(2014)arXiv_06_4271} and the CORNISH UCHII catalog \citep{Hoare(2012)PASP_124_939,Purcell(2013)ApJS_205_1} were employed for the search. 
The UWIFE survey is an imaging survey of the [\astFe{II}] 1.644 \um{} emission line, and the CORNISH survey is a 5 GHz radio continuum survey, both of which cover the 1st Galactic quadrant.
The typical seeings of the two surveys are $\sim0.8''$ (UWIFE) and $\sim1.5''$ (CORNISH), respectively.

Five out of 237 UCHIIs showed nearby [\astFe{II}] features, one of which was less plausible and was tagged as a candidate.
We {propose} that these [\astFe{II}] features might be the ``footprint'' outflow features we are searching for, but more detailed and targeted study is required to clarify it. 
This is based on the following observational facts and estimations: the morphological relation between the [\astFe{II}] and 5 GHz radio features, \Mout{} estimated from the [\astFe{II}] flux, the travel time of the [\astFe{II}] features, and the existence of  several YSO candidates near the UCHIIs.
The UCHIIs accompanying the [\astFe{II}] features have a relatively higher peak flux density in 5 GHz.
\del{\Mout{} shows no significant correlation with the peak flux density of UCHIIs, being limited by small number of data points.}
The fraction of UCHIIs accompanying the [\astFe{II}] features, 5/237, is small compared to the $\sim90$ \% detection rate of high-velocity CO gas around UCHIIs.
We discuss the reasons for this low detection rate, including \mod{the extinction, }the configuration of the outflow cavity, the weakness of [\astFe{II}] features, {the low outflow velocity}, and the outflowing type.
\mod{We note that the sub-arcsec interferometric observations of CO rotational lines would be useful in studying the ``footprint'' outflow features, considering its high detection rate.}

\acknowledgments
J.-H. S. is grateful to Melvin Hoare and Stuart Lumsden for the discussion on the results.
H.-J. K. was supported by NRF(National Research Foundation of Korea) Grant funded by the Korean Government (NRF-2012-Fostering Core Leaders of the Future Basic Science Program).

\bibliographystyle{../_bibtex/bst/apj}
\bibliography{../_bibtex/paper,../_bibtex/book,../_bibtex/manual}

\begin{thebibliography}{96}
\expandafter\ifx\csname natexlab\endcsname\relax\def\natexlab#1{#1}\fi

\bibitem[{{Acord} {et~al.}(1998){Acord}, {Churchwell}, \&
  {Wood}}]{Acord(1998)ApJ_495_L107}
{Acord}, J.~M., {Churchwell}, E., \& {Wood}, D.~O.~S. 1998, ApJ, 495, L107

\bibitem[{{Allen} {et~al.}(2008){Allen}, {Groves}, {Dopita}, {Sutherland}, \&
  {Kewley}}]{Allen(2008)ApJS_178_20}
{Allen}, M.~G., {Groves}, B.~A., {Dopita}, M.~A., {Sutherland}, R.~S., \&
  {Kewley}, L.~J. 2008, ApJS, 178, 20

\bibitem[{{Alonso-Herrero} {et~al.}(1997){Alonso-Herrero}, {Rieke}, {Rieke}, \&
  {Ruiz}}]{Alonso-Herrero(1997)ApJ_482_747}
{Alonso-Herrero}, A., {Rieke}, M.~J., {Rieke}, G.~H., \& {Ruiz}, M. 1997, ApJ,
  482, 747

\bibitem[{{Anderson} \& {Bania}(2009)}]{Anderson(2009)ApJ_690_706}
{Anderson}, L.~D., \& {Bania}, T.~M. 2009, ApJ, 690, 706

\bibitem[{{Antoniucci} {et~al.}(2014){Antoniucci}, {La Camera}, {Nisini},
  {Giannini}, {Lorenzetti}, {Paris}, \& {Sani}}]{Antoniucci(2014)A&A_566_A129}
{Antoniucci}, S., {La Camera}, A., {Nisini}, B., {et~al.} 2014, A\&A, 566, A129

\bibitem[{{Araya} {et~al.}(2002){Araya}, {Hofner}, {Churchwell}, \&
  {Kurtz}}]{Araya(2002)ApJS_138_63}
{Araya}, E., {Hofner}, P., {Churchwell}, E., \& {Kurtz}, S. 2002, ApJS, 138, 63

\bibitem[{{Arce} {et~al.}(2007){Arce}, {Shepherd}, {Gueth}, {Lee}, {Bachiller},
  {Rosen}, \& {Beuther}}]{Arce(2007)2007prpl.conf__245}
{Arce}, H.~G., {Shepherd}, D., {Gueth}, F., {et~al.} 2007, Protostars and
  Planets V, 245

\bibitem[{{Asplund} {et~al.}(2009){Asplund}, {Grevesse}, {Sauval}, \&
  {Scott}}]{Asplund(2009)ARA&A_47_481}
{Asplund}, M., {Grevesse}, N., {Sauval}, A.~J., \& {Scott}, P. 2009, ARA\&A,
  47, 481

\bibitem[{{Bachiller}(1996)}]{Bachiller(1996)ARA&A_34_111}
{Bachiller}, R. 1996, ARA\&A, 34, 111

\bibitem[{{Beck-Winchatz} {et~al.}(1996){Beck-Winchatz}, {Bohm}, \&
  {Noriega-Crespo}}]{Beck-Winchatz(1996)AJ_111_346}
{Beck-Winchatz}, B., {Bohm}, K.-H., \& {Noriega-Crespo}, A. 1996, AJ, 111, 346

\bibitem[{{Bergin} \& {Tafalla}(2007)}]{Bergin(2007)ARA&A_45_339}
{Bergin}, E.~A., \& {Tafalla}, M. 2007, ARA\&A, 45, 339

\bibitem[{{Bessell} \& {Brett}(1988)}]{Bessell(1988)PASP_100_1134}
{Bessell}, M.~S., \& {Brett}, J.~M. 1988, PASP, 100, 1134

\bibitem[{{Beuther} {et~al.}(2002){Beuther}, {Schilke}, {Sridharan}, {Menten},
  {Walmsley}, \& {Wyrowski}}]{Beuther(2002)A&A_383_892}
{Beuther}, H., {Schilke}, P., {Sridharan}, T.~K., {et~al.} 2002, A\&A, 383, 892

\bibitem[{{Blum} {et~al.}(2000){Blum}, {Conti}, \&
  {Damineli}}]{Blum(2000)AJ_119_1860}
{Blum}, R.~D., {Conti}, P.~S., \& {Damineli}, A. 2000, AJ, 119, 1860

\bibitem[{{Burton} {et~al.}(1990){Burton}, {Hollenbach}, \&
  {Tielens}}]{Burton(1990)ApJ_365_620}
{Burton}, M.~G., {Hollenbach}, D.~J., \& {Tielens}, A.~G.~G.~M. 1990, ApJ, 365,
  620

\bibitem[{{Carpenter}(2001)}]{Carpenter(2001)AJ_121_2851}
{Carpenter}, J.~M. 2001, AJ, 121, 2851

\bibitem[{{Casali} {et~al.}(2007){Casali}, {Adamson}, {Alves de Oliveira},
  {Almaini}, {Burch}, {Chuter}, {Elliot}, {Folger}, {Foucaud}, {Hambly},
  {Hastie}, {Henry}, {Hirst}, {Irwin}, {Ives}, {Lawrence}, {Laidlaw}, {Lee},
  {Lewis}, {Lunney}, {McLay}, {Montgomery}, {Pickup}, {Read}, {Rees}, {Robson},
  {Sekiguchi}, {Vick}, {Warren}, \& {Woodward}}]{Casali(2007)A&A_467_777}
{Casali}, M., {Adamson}, A., {Alves de Oliveira}, C., {et~al.} 2007, A\&A, 467,
  777

\bibitem[{{Churchwell}(2002)}]{Churchwell(2002)ARA&A_40_27}
{Churchwell}, E. 2002, ARA\&A, 40, 27

\bibitem[{{Churchwell} {et~al.}(2010){Churchwell}, {Sievers}, \&
  {Thum}}]{Churchwell(2010)A&A_513_A9}
{Churchwell}, E., {Sievers}, A., \& {Thum}, C. 2010, A\&A, 513, A9

\bibitem[{{Churchwell} {et~al.}(1990){Churchwell}, {Walmsley}, \&
  {Cesaroni}}]{Churchwell(1990)A&AS_83_119}
{Churchwell}, E., {Walmsley}, C.~M., \& {Cesaroni}, R. 1990, A\&AS, 83, 119

\bibitem[{{Cohen} {et~al.}(2003){Cohen}, {Wheaton}, \&
  {Megeath}}]{Cohen(2003)AJ_126_1090}
{Cohen}, M., {Wheaton}, W.~A., \& {Megeath}, S.~T. 2003, AJ, 126, 1090

\bibitem[{{Cooper} {et~al.}(2013){Cooper}, {Lumsden}, {Oudmaijer}, {Hoare},
  {Clarke}, {Urquhart}, {Mottram}, {Moore}, \&
  {Davies}}]{Cooper(2013)MNRAS_430_1125}
{Cooper}, H.~D.~B., {Lumsden}, S.~L., {Oudmaijer}, R.~D., {et~al.} 2013, MNRAS,
  430, 1125

\bibitem[{{Cyganowski} {et~al.}(2009){Cyganowski}, {Brogan}, {Hunter}, \&
  {Churchwell}}]{Cyganowski(2009)ApJ_702_1615}
{Cyganowski}, C.~J., {Brogan}, C.~L., {Hunter}, T.~R., \& {Churchwell}, E.
  2009, ApJ, 702, 1615

\bibitem[{{Cyganowski} {et~al.}(2012){Cyganowski}, {Brogan}, {Hunter}, {Zhang},
  {Friesen}, {Indebetouw}, \& {Chandler}}]{Cyganowski(2012)ApJ_760_L20}
{Cyganowski}, C.~J., {Brogan}, C.~L., {Hunter}, T.~R., {et~al.} 2012, ApJ, 760,
  L20

\bibitem[{{Downes} {et~al.}(1980){Downes}, {Wilson}, {Bieging}, \&
  {Wink}}]{Downes(1980)A&AS_40_379}
{Downes}, D., {Wilson}, T.~L., {Bieging}, J., \& {Wink}, J. 1980, A\&AS, 40,
  379

\bibitem[{{Draine}(2003)}]{Draine(2003)ARA&A_41_241}
{Draine}, B.~T. 2003, ARA\&A, 41, 241

\bibitem[{{Draine} \& {McKee}(1993)}]{Draine(1993)ARA&A_31_373}
{Draine}, B.~T., \& {McKee}, C.~F. 1993, ARA\&A, 31, 373

\bibitem[{{Dye} {et~al.}(2006){Dye}, {Warren}, {Hambly}, {Cross}, {Hodgkin},
  {Irwin}, {Lawrence}, {Adamson}, {Almaini}, {Edge}, {Hirst}, {Jameson},
  {Lucas}, {van Breukelen}, {Bryant}, {Casali}, {Collins}, {Dalton}, {Davies},
  {Davis}, {Emerson}, {Evans}, {Foucaud}, {Gonzales-Solares}, {Hewett},
  {Kendall}, {Kerr}, {Leggett}, {Lodieu}, {Loveday}, {Lewis}, {Mann},
  {McMahon}, {Mortlock}, {Nakajima}, {Pinfield}, {Rawlings}, {Read}, {Riello},
  {Sekiguchi}, {Smith}, {Sutorius}, {Varricatt}, {Walton}, \&
  {Weatherley}}]{Dye(2006)MNRAS_372_1227}
{Dye}, S., {Warren}, S.~J., {Hambly}, N.~C., {et~al.} 2006, MNRAS, 372, 1227

\bibitem[{{Ellerbroek} {et~al.}(2013){Ellerbroek}, {Podio}, {Kaper}, {Sana},
  {Huppenkothen}, {de Koter}, \& {Monaco}}]{Ellerbroek(2013)A&A_551_A5}
{Ellerbroek}, L.~E., {Podio}, L., {Kaper}, L., {et~al.} 2013, A\&A, 551, A5

\bibitem[{{Flower} \& {Pineau Des
  For{\^e}ts}(2010)}]{Flower(2010)MNRAS_406_1745}
{Flower}, D.~R., \& {Pineau Des For{\^e}ts}, G. 2010, MNRAS, 406, 1745

\bibitem[{{Fox} {et~al.}(2009){Fox}, {Skrutskie}, {Chevalier}, {Kanneganti},
  {Park}, {Wilson}, {Nelson}, {Amirhadji}, {Crump}, {Hoeft}, {Provence},
  {Sargeant}, {Sop}, {Tea}, {Thomas}, \& {Woolard}}]{Fox(2009)ApJ_691_650}
{Fox}, O., {Skrutskie}, M.~F., {Chevalier}, R.~A., {et~al.} 2009, ApJ, 691, 650

\bibitem[{{Franco-Hern{\'a}ndez} \&
  {Rodr{\'{\i}}guez}(2004)}]{Franco-Hernandez(2004)ApJ_604_L105}
{Franco-Hern{\'a}ndez}, R., \& {Rodr{\'{\i}}guez}, L.~F. 2004, ApJ, 604, L105

\bibitem[{{Frank} {et~al.}(2014){Frank}, {Ray}, {Cabrit}, {Hartigan}, {Arce},
  {Bacciotti}, {Bally}, {Benisty}, {Eisl{\"o}ffel}, {G{\"u}del}, {Lebedev},
  {Nisini}, \& {Raga}}]{Frank(2014)arXiv_02_3553}
{Frank}, A., {Ray}, T.~P., {Cabrit}, S., {et~al.} 2014, arXiv, 02, 3553

\bibitem[{{Froebrich} {et~al.}(2011){Froebrich}, {Davis}, {Ioannidis},
  {Gledhill}, {Takami}, {Chrysostomou}, {Drew}, {Eisl{\"o}ffel}, {Gosling},
  {Gredel}, {Hatchell}, {Hodapp}, {Kumar}, {Lucas}, {Matthews}, {Rawlings},
  {Smith}, {Stecklum}, {Varricatt}, {Lee}, {Teixeira}, {Aspin}, {Khanzadyan},
  {Karr}, {Kim}, {Koo}, {Lee}, {Lee}, {Magakian}, {Movsessian}, {Nikogossian},
  {Pyo}, \& {Stanke}}]{Froebrich(2011)MNRAS_413_480}
{Froebrich}, D., {Davis}, C.~J., {Ioannidis}, G., {et~al.} 2011, MNRAS, 413,
  480

\bibitem[{{Galv{\'a}n-Madrid} {et~al.}(2008){Galv{\'a}n-Madrid},
  {Rodr{\'{\i}}guez}, {Ho}, \& {Keto}}]{Galvan-Madrid(2008)ApJ_674_L33}
{Galv{\'a}n-Madrid}, R., {Rodr{\'{\i}}guez}, L.~F., {Ho}, P.~T.~P., \& {Keto},
  E. 2008, ApJ, 674, L33

\bibitem[{{Giannini} {et~al.}(2008){Giannini}, {Calzoletti}, {Nisini}, {Davis},
  {Eisl{\"o}ffel}, \& {Smith}}]{Giannini(2008)A&A_481_123}
{Giannini}, T., {Calzoletti}, L., {Nisini}, B., {et~al.} 2008, A\&A, 481, 123

\bibitem[{{Giannini} {et~al.}(2013){Giannini}, {Nisini}, {Antoniucci},
  {Alcal{\'a}}, {Bacciotti}, {Bonito}, {Podio}, {Stelzer}, \&
  {Whelan}}]{Giannini(2013)ApJ_778_71}
{Giannini}, T., {Nisini}, B., {Antoniucci}, S., {et~al.} 2013, ApJ, 778, 71

\bibitem[{{Gonz{\'a}lez-Avil{\'e}s} {et~al.}(2005){Gonz{\'a}lez-Avil{\'e}s},
  {Lizano}, \& {Raga}}]{Gonzalez-Aviles(2005)ApJ_621_359}
{Gonz{\'a}lez-Avil{\'e}s}, M., {Lizano}, S., \& {Raga}, A.~C. 2005, ApJ, 621,
  359

\bibitem[{{Guzm{\'a}n} {et~al.}(2012){Guzm{\'a}n}, {Garay}, {Brooks}, \&
  {Voronkov}}]{Guzman(2012)ApJ_753_51}
{Guzm{\'a}n}, A.~E., {Garay}, G., {Brooks}, K.~J., \& {Voronkov}, M.~A. 2012,
  ApJ, 753, 51

\bibitem[{{Hanson} {et~al.}(2002){Hanson}, {Luhman}, \&
  {Rieke}}]{Hanson(2002)ApJS_138_35}
{Hanson}, M.~M., {Luhman}, K.~L., \& {Rieke}, G.~H. 2002, ApJS, 138, 35

\bibitem[{{Hartmann} \& {Kenyon}(1996)}]{Hartmann(1996)ARA&A_34_207}
{Hartmann}, L., \& {Kenyon}, S.~J. 1996, ARA\&A, 34, 207

\bibitem[{{Hewett} {et~al.}(2006){Hewett}, {Warren}, {Leggett}, \&
  {Hodgkin}}]{Hewett(2006)MNRAS_367_454}
{Hewett}, P.~C., {Warren}, S.~J., {Leggett}, S.~K., \& {Hodgkin}, S.~T. 2006,
  MNRAS, 367, 454

\bibitem[{{Hoare} {et~al.}(2007){Hoare}, {Kurtz}, {Lizano}, {Keto}, \&
  {Hofner}}]{Hoare(2007)2007prpl.conf__181}
{Hoare}, M.~G., {Kurtz}, S.~E., {Lizano}, S., {Keto}, E., \& {Hofner}, P. 2007,
  2007prpl.conf, 181

\bibitem[{{Hoare} {et~al.}(2012){Hoare}, {Purcell}, {Churchwell}, {Diamond},
  {Cotton}, {Chandler}, {Smethurst}, {Kurtz}, {Mundy}, {Dougherty}, {Fender},
  {Fuller}, {Jackson}, {Garrington}, {Gledhill}, {Goldsmith}, {Lumsden},
  {Mart{\'{\i}}}, {Moore}, {Muxlow}, {Oudmaijer}, {Pandian}, {Paredes},
  {Shepherd}, {Spencer}, {Thompson}, {Umana}, {Urquhart}, \&
  {Zijlstra}}]{Hoare(2012)PASP_124_939}
{Hoare}, M.~G., {Purcell}, C.~R., {Churchwell}, E.~B., {et~al.} 2012, PASP,
  124, 939

\bibitem[{{Hodgkin} {et~al.}(2009){Hodgkin}, {Irwin}, {Hewett}, \&
  {Warren}}]{Hodgkin(2009)MNRAS_394_675}
{Hodgkin}, S.~T., {Irwin}, M.~J., {Hewett}, P.~C., \& {Warren}, S.~J. 2009,
  MNRAS, 394, 675

\bibitem[{{Hollenbach} \& {McKee}(1989)}]{Hollenbach(1989)ApJ_342_306}
{Hollenbach}, D., \& {McKee}, C.~F. 1989, ApJ, 342, 306

\bibitem[{{Hollenbach} \& {Tielens}(1997)}]{Hollenbach(1997)ARA&A_35_179}
{Hollenbach}, D.~J., \& {Tielens}, A.~G.~G.~M. 1997, ARA\&A, 35, 179

\bibitem[{{Jones} \& {Dickey}(2012)}]{Jones(2012)ApJ_753_62}
{Jones}, C., \& {Dickey}, J.~M. 2012, ApJ, 753, 62

\bibitem[{{Kaufman} \& {Neufeld}(1996)}]{Kaufman(1996)ApJ_456_611}
{Kaufman}, M.~J., \& {Neufeld}, D.~A. 1996, ApJ, 456, 611

\bibitem[{{Keto}(2002)}]{Keto(2002)ApJ_568_754}
{Keto}, E. 2002, ApJ, 568, 754

\bibitem[{{Keto}(2003)}]{Keto(2003)ApJ_599_1196}
---. 2003, ApJ, 599, 1196

\bibitem[{{Keto} \& {Klaassen}(2008)}]{Keto(2008)ApJ_678_L109}
{Keto}, E., \& {Klaassen}, P. 2008, ApJ, 678, L109

\bibitem[{{Keto} \& {Wood}(2006)}]{Keto(2006)ApJ_637_850}
{Keto}, E., \& {Wood}, K. 2006, ApJ, 637, 850

\bibitem[{{Kim} \& {Koo}(2001)}]{Kim(2001)ApJ_549_979}
{Kim}, K., \& {Koo}, B. 2001, ApJ, 549, 979

\bibitem[{{Kolpak} {et~al.}(2003){Kolpak}, {Jackson}, {Bania}, {Clemens}, \&
  {Dickey}}]{Kolpak(2003)ApJ_582_756}
{Kolpak}, M.~A., {Jackson}, J.~M., {Bania}, T.~M., {Clemens}, D.~P., \&
  {Dickey}, J.~M. 2003, ApJ, 582, 756

\bibitem[{{Koornneef}(1983)}]{Koornneef(1983)A&A_128_84}
{Koornneef}, J. 1983, A\&A, 128, 84

\bibitem[{{Krumholz} {et~al.}(2012){Krumholz}, {Klein}, \&
  {McKee}}]{Krumholz(2012)ApJ_754_71}
{Krumholz}, M.~R., {Klein}, R.~I., \& {McKee}, C.~F. 2012, ApJ, 754, 71

\bibitem[{{Krumholz} {et~al.}(2009){Krumholz}, {Klein}, {McKee}, {Offner}, \&
  {Cunningham}}]{Krumholz(2009)Science_323_754}
{Krumholz}, M.~R., {Klein}, R.~I., {McKee}, C.~F., {Offner}, S.~S.~R., \&
  {Cunningham}, A.~J. 2009, Science, 323, 754

\bibitem[{{Kuiper} \& {Yorke}(2013{\natexlab{a}})}]{Kuiper(2013)ApJ_763_104}
{Kuiper}, R., \& {Yorke}, H.~W. 2013{\natexlab{a}}, ApJ, 763, 104

\bibitem[{{Kuiper} \& {Yorke}(2013{\natexlab{b}})}]{Kuiper(2013)ApJ_772_61}
---. 2013{\natexlab{b}}, ApJ, 772, 61

\bibitem[{{Kurtz} {et~al.}(1994){Kurtz}, {Churchwell}, \&
  {Wood}}]{Kurtz(1994)ApJS_91_659}
{Kurtz}, S., {Churchwell}, E., \& {Wood}, D.~O.~S. 1994, ApJS, 91, 659

\bibitem[{{Lada}(1985)}]{Lada(1985)ARA&A_23_267}
{Lada}, C.~J. 1985, ARA\&A, 23, 267

\bibitem[{{Lawrence} {et~al.}(2007){Lawrence}, {Warren}, {Almaini}, {Edge},
  {Hambly}, {Jameson}, {Lucas}, {Casali}, {Adamson}, {Dye}, {Emerson},
  {Foucaud}, {Hewett}, {Hirst}, {Hodgkin}, {Irwin}, {Lodieu}, {McMahon},
  {Simpson}, {Smail}, {Mortlock}, \& {Folger}}]{Lawrence(2007)MNRAS_379_1599}
{Lawrence}, A., {Warren}, S.~J., {Almaini}, O., {et~al.} 2007, MNRAS, 379, 1599

\bibitem[{{Le Bourlot} {et~al.}(2002){Le Bourlot}, {Pineau des For{\^e}ts},
  {Flower}, \& {Cabrit}}]{LeBourlot(2002)MNRAS_332_985}
{Le Bourlot}, J., {Pineau des For{\^e}ts}, G., {Flower}, D.~R., \& {Cabrit}, S.
  2002, MNRAS, 332, 985

\bibitem[{{Lee} {et~al.}(2014){Lee}, {Koo}, {Lee}, {Lee}, {Shinn}, {Kim},
  {Kim}, {Pyo}, {Moon}, {Yoon}, {Chun}, {Froebrich}, {Davis}, {Varricatt},
  {Kyeong}, {Hwang}, {Park}, {Lee}, {Lee}, \&
  {Ishiguro}}]{Lee(2014)arXiv_06_4271}
{Lee}, J.-J., {Koo}, B.-C., {Lee}, Y.-H., {et~al.} 2014, arXiv, 06, 4271,
  accepted in MNRAS

\bibitem[{{Lucas} {et~al.}(2008){Lucas}, {Hoare}, {Longmore}, {Schr{\"o}der},
  {Davis}, {Adamson}, {Bandyopadhyay}, {de Grijs}, {Smith}, {Gosling},
  {Mitchison}, {G{\'a}sp{\'a}r}, {Coe}, {Tamura}, {Parker}, {Irwin}, {Hambly},
  {Bryant}, {Collins}, {Cross}, {Evans}, {Gonzalez-Solares}, {Hodgkin},
  {Lewis}, {Read}, {Riello}, {Sutorius}, {Lawrence}, {Drew}, {Dye}, \&
  {Thompson}}]{Lucas(2008)MNRAS_391_136}
{Lucas}, P.~W., {Hoare}, M.~G., {Longmore}, A., {et~al.} 2008, MNRAS, 391, 136

\bibitem[{{Mois{\'e}s} {et~al.}(2011){Mois{\'e}s}, {Damineli}, {Figuer{\^e}do},
  {Blum}, {Conti}, \& {Barbosa}}]{Moises(2011)MNRAS_411_705}
{Mois{\'e}s}, A.~P., {Damineli}, A., {Figuer{\^e}do}, E., {et~al.} 2011, MNRAS,
  411, 705

\bibitem[{{Mottram} {et~al.}(2011){Mottram}, {Hoare}, {Davies}, {Lumsden},
  {Oudmaijer}, {Urquhart}, {Moore}, {Cooper}, \&
  {Stead}}]{Mottram(2011)ApJ_730_L33}
{Mottram}, J.~C., {Hoare}, M.~G., {Davies}, B., {et~al.} 2011, ApJ, 730, L33

\bibitem[{{Mouri} \& {Taniguchi}(2000)}]{Mouri(2000)ApJ_534_L63}
{Mouri}, H., \& {Taniguchi}, Y. 2000, ApJ, 534, L63

\bibitem[{{Neufeld} \& {Dalgarno}(1989)}]{Neufeld(1989)ApJ_344_251}
{Neufeld}, D.~A., \& {Dalgarno}, A. 1989, ApJ, 344, 251

\bibitem[{{Nisini} {et~al.}(2005){Nisini}, {Bacciotti}, {Giannini}, {Massi},
  {Eisl{\"o}ffel}, {Podio}, \& {Ray}}]{Nisini(2005)A&A_441_159}
{Nisini}, B., {Bacciotti}, F., {Giannini}, T., {et~al.} 2005, A\&A, 441, 159

\bibitem[{{Nisini} {et~al.}(2002){Nisini}, {Caratti o Garatti}, {Giannini}, \&
  {Lorenzetti}}]{Nisini(2002)A&A_393_1035}
{Nisini}, B., {Caratti o Garatti}, A., {Giannini}, T., \& {Lorenzetti}, D.
  2002, A\&A, 393, 1035

\bibitem[{{Podio} {et~al.}(2006){Podio}, {Bacciotti}, {Nisini},
  {Eisl{\"o}ffel}, {Massi}, {Giannini}, \& {Ray}}]{Podio(2006)A&A_456_189}
{Podio}, L., {Bacciotti}, F., {Nisini}, B., {et~al.} 2006, A\&A, 456, 189

\bibitem[{{Purcell} {et~al.}(2013){Purcell}, {Hoare}, {Cotton}, {Lumsden},
  {Urquhart}, {Chandler}, {Churchwell}, {Diamond}, {Dougherty}, {Fender},
  {Fuller}, {Garrington}, {Gledhill}, {Goldsmith}, {Hindson}, {Jackson},
  {Kurtz}, {Mart{\'{\i}}}, {Moore}, {Mundy}, {Muxlow}, {Oudmaijer}, {Pandian},
  {Paredes}, {Shepherd}, {Smethurst}, {Spencer}, {Thompson}, {Umana}, \&
  {Zijlstra}}]{Purcell(2013)ApJS_205_1}
{Purcell}, C.~R., {Hoare}, M.~G., {Cotton}, W.~D., {et~al.} 2013, ApJS, 205, 1

\bibitem[{{Radhakrishnan} {et~al.}(1972){Radhakrishnan}, {Goss}, {Murray}, \&
  {Brooks}}]{Radhakrishnan(1972)ApJS_24_49}
{Radhakrishnan}, V., {Goss}, W.~M., {Murray}, J.~D., \& {Brooks}, J.~W. 1972,
  ApJS, 24, 49

\bibitem[{{Ray} {et~al.}(2007){Ray}, {Dougados}, {Bacciotti}, {Eisl{\"o}ffel},
  \& {Chrysostomou}}]{Ray(2007)2007prpl.conf__231}
{Ray}, T., {Dougados}, C., {Bacciotti}, F., {Eisl{\"o}ffel}, J., \&
  {Chrysostomou}, A. 2007, 2007prpl.conf, 231

\bibitem[{{Reipurth} \& {Bally}(2001)}]{Reipurth(2001)ARA&A_39_403}
{Reipurth}, B., \& {Bally}, J. 2001, ARA\&A, 39, 403

\bibitem[{{Rieke} \& {Lebofsky}(1985)}]{Rieke(1985)ApJ_288_618}
{Rieke}, G.~H., \& {Lebofsky}, M.~J. 1985, ApJ, 288, 618

\bibitem[{{San Jos{\'e}-Garc{\'{\i}}a} {et~al.}(2013){San
  Jos{\'e}-Garc{\'{\i}}a}, {Mottram}, {Kristensen}, {van Dishoeck},
  {Y{\i}ld{\i}z}, {van der Tak}, {Herpin}, {Visser}, {McCoey}, {Wyrowski},
  {Braine}, \& {Johnstone}}]{SanJose-Garcia(2013)A&A_553_A125}
{San Jos{\'e}-Garc{\'{\i}}a}, I., {Mottram}, J.~C., {Kristensen}, L.~E.,
  {et~al.} 2013, A\&A, 553, A125

\bibitem[{{Shepherd} \& {Churchwell}(1996)}]{Shepherd(1996)ApJ_457_267}
{Shepherd}, D.~S., \& {Churchwell}, E. 1996, ApJ, 457, 267

\bibitem[{{Shinn} {et~al.}(2013){Shinn}, {Pyo}, {Lee}, {Lee}, {Kim}, {Koo},
  {Sung}, {Chun}, {Lyo}, {Moon}, {Kyeong}, {Park}, {Hur}, \&
  {Lee}}]{Shinn(2013)ApJ_777_45}
{Shinn}, J.-H., {Pyo}, T.-S., {Lee}, J.-J., {et~al.} 2013, ApJ, 777, 45

\bibitem[{{Skrutskie} {et~al.}(2006){Skrutskie}, {Cutri}, {Stiening},
  {Weinberg}, {Schneider}, {Carpenter}, {Beichman}, {Capps}, {Chester},
  {Elias}, {Huchra}, {Liebert}, {Lonsdale}, {Monet}, {Price}, {Seitzer},
  {Jarrett}, {Kirkpatrick}, {Gizis}, {Howard}, {Evans}, {Fowler}, {Fullmer},
  {Hurt}, {Light}, {Kopan}, {Marsh}, {McCallon}, {Tam}, {Van Dyk}, \&
  {Wheelock}}]{Skrutskie(2006)AJ_131_1163}
{Skrutskie}, M.~F., {Cutri}, R.~M., {Stiening}, R., {et~al.} 2006, AJ, 131,
  1163

\bibitem[{{Thompson} {et~al.}(2006){Thompson}, {Hatchell}, {Walsh},
  {MacDonald}, \& {Millar}}]{Thompson(2006)A&A_453_1003}
{Thompson}, M.~A., {Hatchell}, J., {Walsh}, A.~J., {MacDonald}, G.~H., \&
  {Millar}, T.~J. 2006, A\&A, 453, 1003

\bibitem[{{van der Tak} {et~al.}(2005){van der Tak}, {Tuthill}, \&
  {Danchi}}]{vanderTak(2005)A&A_431_993}
{van der Tak}, F.~F.~S., {Tuthill}, P.~G., \& {Danchi}, W.~C. 2005, A\&A, 431,
  993

\bibitem[{{Varricatt} {et~al.}(2010){Varricatt}, {Davis}, {Ramsay}, \&
  {Todd}}]{Varricatt(2010)MNRAS_404_661}
{Varricatt}, W.~P., {Davis}, C.~J., {Ramsay}, S., \& {Todd}, S.~P. 2010, MNRAS,
  404, 661

\bibitem[{{Walsh} {et~al.}(1998){Walsh}, {Burton}, {Hyland}, \&
  {Robinson}}]{Walsh(1998)MNRAS_301_640}
{Walsh}, A.~J., {Burton}, M.~G., {Hyland}, A.~R., \& {Robinson}, G. 1998,
  MNRAS, 301, 640

\bibitem[{{Waters} \& {Waelkens}(1998)}]{Waters(1998)ARA&A_36_233}
{Waters}, L.~B.~F.~M., \& {Waelkens}, C. 1998, ARA\&A, 36, 233

\bibitem[{{Watson} {et~al.}(2003){Watson}, {Araya}, {Sewilo}, {Churchwell},
  {Hofner}, \& {Kurtz}}]{Watson(2003)ApJ_587_714}
{Watson}, C., {Araya}, E., {Sewilo}, M., {et~al.} 2003, ApJ, 587, 714

\bibitem[{{Watson} {et~al.}(2007){Watson}, {Churchwell}, {Zweibel}, \&
  {Crutcher}}]{Watson(2007)ApJ_657_318}
{Watson}, C., {Churchwell}, E., {Zweibel}, E.~G., \& {Crutcher}, R.~M. 2007,
  ApJ, 657, 318

\bibitem[{{Weingartner} \& {Draine}(2001)}]{Weingartner(2001)ApJ_548_296}
{Weingartner}, J.~C., \& {Draine}, B.~T. 2001, ApJ, 548, 296

\bibitem[{{Wilgenbus} {et~al.}(2000){Wilgenbus}, {Cabrit}, {Pineau des
  For{\^e}ts}, \& {Flower}}]{Wilgenbus(2000)A&A_356_1010}
{Wilgenbus}, D., {Cabrit}, S., {Pineau des For{\^e}ts}, G., \& {Flower}, D.~R.
  2000, A\&A, 356, 1010

\bibitem[{{Wilson}(1972)}]{Wilson(1972)A&A_19_354}
{Wilson}, T.~L. 1972, A\&A, 19, 354

\bibitem[{{Wood} \& {Churchwell}(1989)}]{Wood(1989)ApJS_69_831}
{Wood}, D.~O.~S., \& {Churchwell}, E. 1989, ApJS, 69, 831

\bibitem[{{Wu} {et~al.}(2004){Wu}, {Wei}, {Zhao}, {Shi}, {Yu}, {Qin}, \&
  {Huang}}]{Wu(2004)A&A_426_503}
{Wu}, Y., {Wei}, Y., {Zhao}, M., {et~al.} 2004, A\&A, 426, 503

\bibitem[{{Yorke} \& {Sonnhalter}(2002)}]{Yorke(2002)ApJ_569_846}
{Yorke}, H.~W., \& {Sonnhalter}, C. 2002, ApJ, 569, 846

\bibitem[{{Zinnecker} \& {Yorke}(2007)}]{Zinnecker(2007)ARA&A_45_481}
{Zinnecker}, H., \& {Yorke}, H.~W. 2007, ARA\&A, 45, 481

\end{thebibliography}
%\bibliographystyle{G:/Work/Publication/bibtex/astronat/apj/apj}
%\bibliography{G:/Work/Publication/bibtex/paper,G:/Work/Publication/bibtex/book,G:/Work/Publication/bibtex/manual} 

% FIGURE & TABLE
%%%%%%%%%% FIGURES
\clearpage

\clearpage
\begin{figure}
%\figurenum{2}
\centering
%\includegraphics[scale=0.9]{f2a.pdf}\\(a) G025.3824-00.1812 (\textit{upper-circle}) and G025.3809-00.1815 (\textit{lower-circle})
%\caption{(Continued)}
\caption{
Outflow features around CORNISH sources seen in the continuum-subtracted [\astFe{II}] (\textit{top panels}) and RGB  composite (\textit{middle panels}; {R=H, G=[\astFe{II}]}, B=H) images. 
The [\astFe{II}] and H images (FWHM $\sim0.8''$) are from the UWIFE \citep{Lee(2014)arXiv_06_4271} and UKIDSS \citep{Lucas(2008)MNRAS_391_136} surveys, respectively.
The CORNISH 5 GHz image (FWHM $\sim1.5''$) is shown for reference in the \textit{bottom-left panel}. 
The \textit{dashed boxes} in the \textit{left panels} indicate the region enlarged in the \textit{right panels}. 
Ellipses indicate where the flux is measured (on-source:\textit{solid}, off-source:\textit{dashed}). 
The ellipse pairs (on+off) are alphabetically tagged in order of increasing RA (cf.~\textit{top-right panel}). 
Circles with a red slash indicate the region excluded during the flux measurement. The measured fluxes are listed in Table \ref{tbl-flux}.
\label{fig-uchii}
}
%\caption{Outflow features around CORNISH sources seen in the continuum-subtracted [\astFe{II}] (\textit{left panels}) and RGB  composite (\textit{right panels};R=[\astFe{II}],G=H,B=H) images. The \textit{dashed boxes} in the \textit{upper panels} indicate the region enlarged in the \textit{lower panels}. Ellipses indicate where the flux is measured (on-source:\textit{solid}, off-source:\textit{dashed}). The ellipse pairs (on+off) is alphabetically tagged in order of increasing RA (cf.~\textit{lower-left} panel). Circles with a red slash indicate the region excluded during the flux measurement. The measured fluxes are listed in Table \ref{tbl-flux}.}
\end{figure}

\clearpage
\begin{figure}
\figurenum{2}
\centering
\includegraphics[scale=0.9]{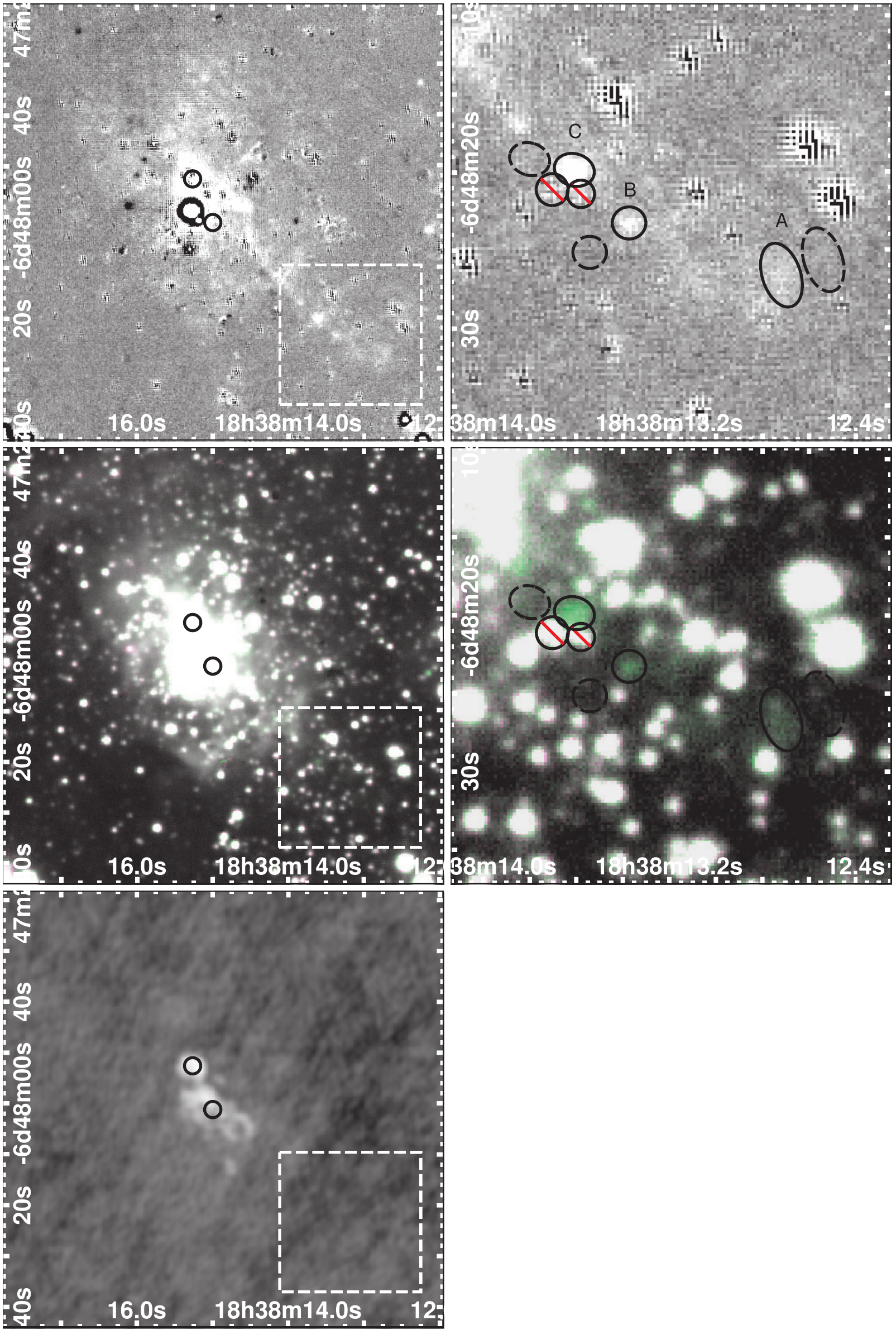}\\(a) G025.3824$-$00.1812 (\textit{upper-circle}) and G025.3809$-$00.1815 (\textit{lower-circle})
\caption{(Continued)}
%\caption{Outflow features around CORNISH sources seen in the continuum-subtracted [\astFe{II}] (\textit{left panels}) and RGB  composite (\textit{right panels};R=[\astFe{II}],G=H,B=H) images. The \textit{dashed boxes} in the \textit{upper panels} indicate the region enlarged in the \textit{lower panels}. Ellipses indicate where the flux is measured (on-source:\textit{solid}, off-source:\textit{dashed}). The ellipse pairs (on+off) is alphabetically tagged in order of increasing RA (cf.~\textit{lower-left} panel). Circles with a red slash indicate the region excluded during the flux measurement. The measured fluxes are listed in Table \ref{tbl-flux}.}
\end{figure}

\clearpage
\begin{figure}
\figurenum{2}
\centering
\includegraphics[scale=0.9]{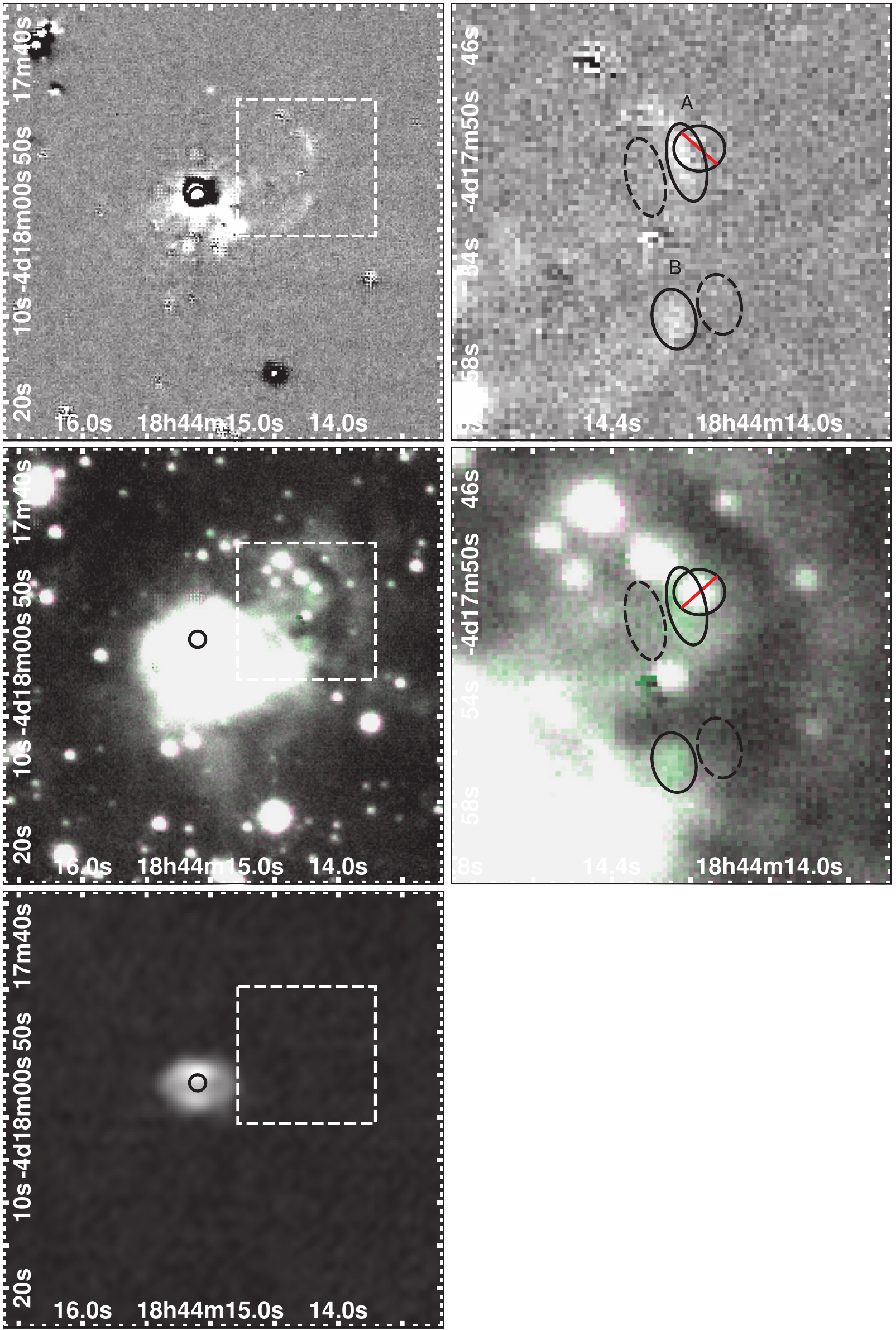}\\(b) G028.2879$-$00.3641
\caption{(Continued)}
\end{figure}

\clearpage
\begin{figure}
\figurenum{2}
\centering
\includegraphics[scale=0.9]{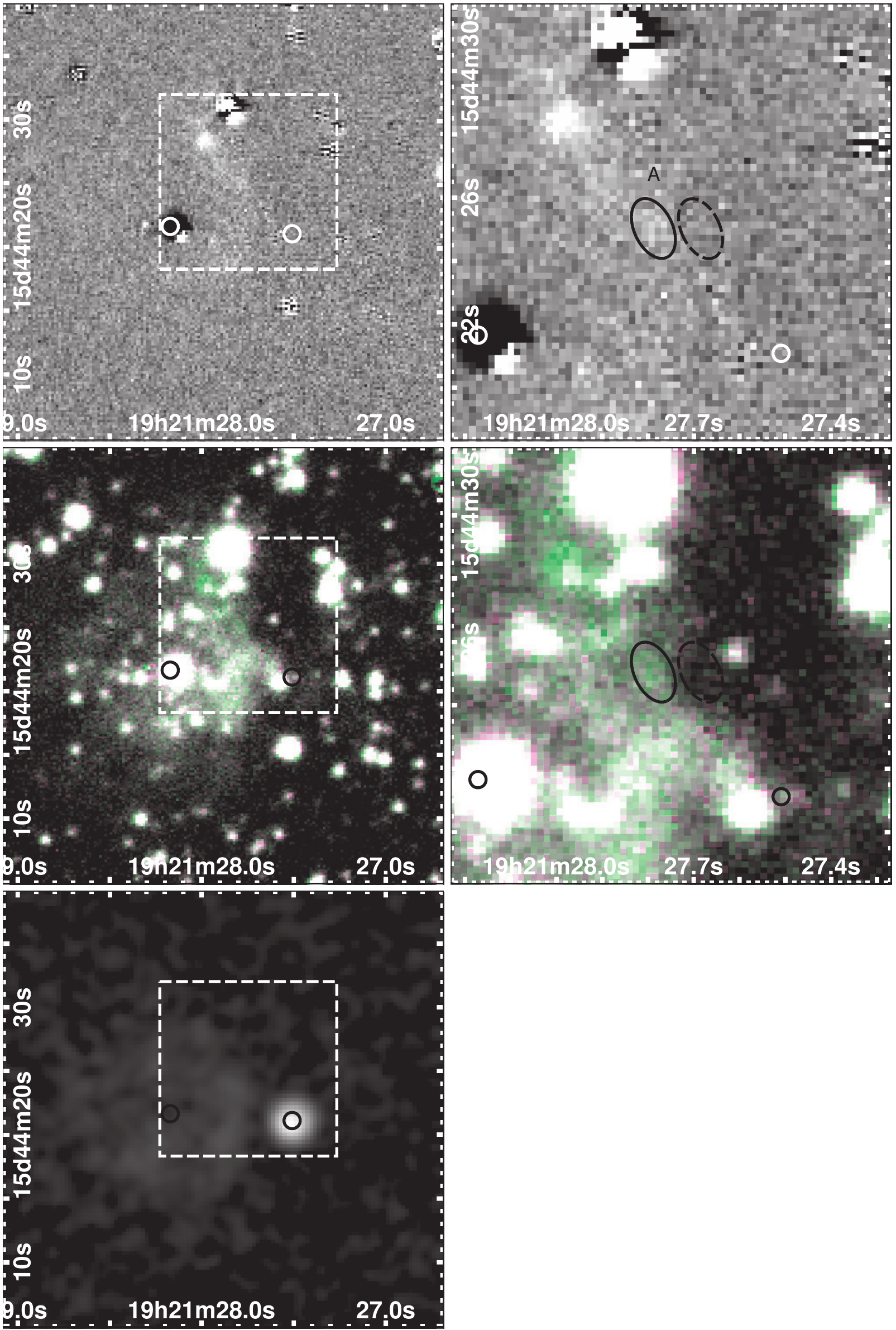}\\(c) G050.3157+00.6747 (\textit{left-circle}) and G050.3152+00.6762 (\textit{right-circle})
\caption{(Continued)}
\end{figure}

\clearpage
\begin{figure}
\centering
\caption{
Candidate outflow features around CORNISH sources G013.8726+00.2818 seen in the continuum-subtracted [\astFe{II}] (\textit{top}) and RGB  composite (\textit{middle};{R=H, G=[\astFe{II}]}, B=H) images. 
%Candidate outflow features around CORNISH sources seen in the continuum-subtracted [\astFe{II}] (\textit{top}) and RGB  composite (\textit{middle};R=[\astFe{II}],G=H,B=H) images. 
The candidate [\astFe{II}] features are indicated by arrows for clarity.
The [\astFe{II}] and H images (FWHM $\sim0.8''$) are from the UWIFE \citep{Lee(2014)arXiv_06_4271} and UKIDSS \citep{Lucas(2008)MNRAS_391_136} surveys, respectively.
The CORNISH 5 GHz image (FWHM $\sim1.5''$) is shown for reference in the \textit{bottom panel}.
\label{fig-uchii-cand}
}
%\caption{Candidate outflow features around CORNISH sources seen in the continuum-subtracted [\astFe{II}] (\textit{left}) and RGB  composite (\textit{right};R=[\astFe{II}],G=H,B=H) images. \label{fig-uchii}}
\end{figure}

\clearpage
\begin{figure}
\figurenum{3}
\centering
\includegraphics[scale=0.9]{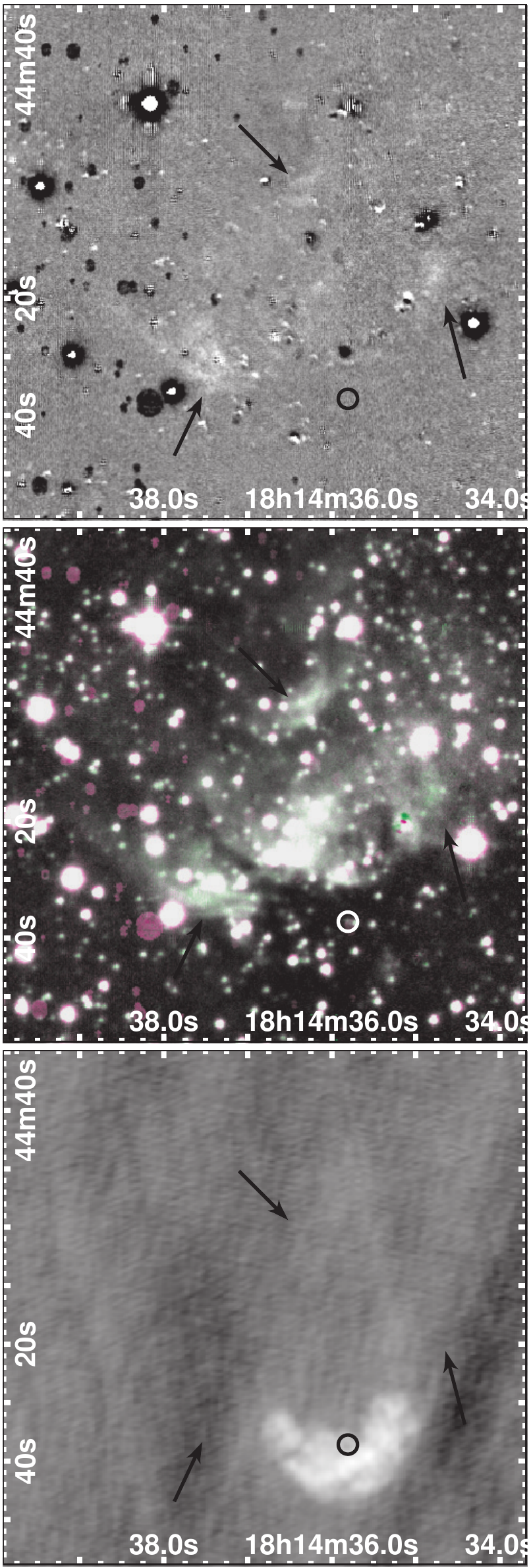}
\caption{(Continued)}
%\caption{Candidate outflow features around CORNISH sources seen in the continuum-subtracted [\astFe{II}] (\textit{left}) and RGB  composite (\textit{right};R=[\astFe{II}],G=H,B=H) images. \label{fig-uchii}}
\end{figure}

%\clearpage
%\begin{figure}
%\figurenum{3}
%\centering
%\includegraphics[scale=0.9]{f3b.pdf}\\(b) G023.9564+00.1493
%\caption{(Continued)}
%\end{figure}
%
%
%
%
%\clearpage
%\begin{figure}
%\figurenum{3}
%\centering
%\includegraphics[scale=0.9]{f3c.pdf}\\(c) G043.1651$-$00.0283 
%\caption{(Continued)}
%\end{figure}
%
%
%\clearpage
%\begin{figure}
%\figurenum{3}
%\centering
%\includegraphics[scale=0.9]{f3d.pdf}\\(d) G045.1242+00.1356 (\textit{upper-circle}) and G045.1223+00.1321 (\textit{lower-circle})
%\caption{(Continued)}
%\end{figure}
%
%
%\clearpage
%\begin{figure}
%\figurenum{3}
%\centering
%\includegraphics[scale=0.9]{f3e.pdf}\\(e) G060.8842$-$00.1286
%\caption{(Continued)}
%\end{figure}
%
%
%\clearpage
%\begin{figure}
%\figurenum{3}
%\centering
%\includegraphics[scale=0.9]{f3f.pdf}\\(f) G061.4770+00.0891 (\textit{upper-circle}) and G061.4763+00.0892 (\textit{lower-circle})
%\caption{(Continued)}
%\end{figure}

\clearpage
\begin{figure}
%\figurenum{3}
\centering
\includegraphics[scale=0.9]{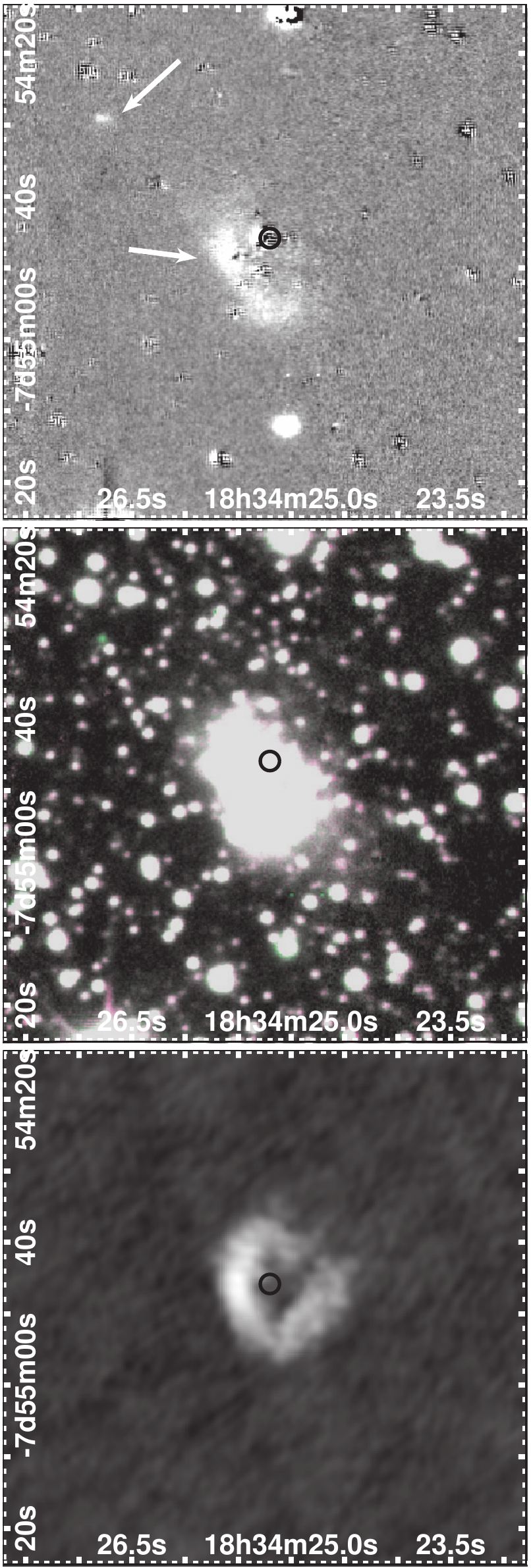}
\caption{The example of excluded [\astFe{II}] features (\textit{arrows}) around UCHII G023.9564+00.1493, seen in the continuum-subtracted [\astFe{II}] image (\textit{top}), RGB  composite image (\textit{middle}; {R=H, G=[\astFe{II}]}, B=H), and CORNISH 5 GHz image (\textit{bottom}). \label{fig-excl}}
\end{figure}

\clearpage
\begin{figure}
%\figurenum{2}
\centering
%\includegraphics[scale=0.9]{f2a.pdf}\\(a) G025.3824-00.1812 (\textit{upper-circle}) and G025.3809-00.1815 (\textit{lower-circle})
%\caption{(Continued)}
\caption{
{
Comparison of [\astFe{II}] features with other near-infrared images: the continuum-subtracted [\astFe{II}] (\textit{top-panels}), the \HtwolineK{} 2.12 \um, K RGB composite image (\textit{middle panels}; R=\Htwo, G=K, B=K), and the J, H, K RGB composite image (\textit{bottom panel}; R=K, G=H, B=J).
The field-of-view setting, boxes, and ellipses are the same as Figure \ref{fig-uchii}.
The \Htwo{} (FWHM $\sim0.7''$) and JHK images (FWHM $\sim0.8''$) are from the UWISH2 \citep{Froebrich(2011)MNRAS_413_480} and UKIDSS \citep{Lucas(2008)MNRAS_391_136} surveys, respectively.
\label{fig-uchii-ukidss}
}
}
%\caption{Outflow features around CORNISH sources seen in the continuum-subtracted [\astFe{II}] (\textit{left panels}) and RGB  composite (\textit{right panels};R=[\astFe{II}],G=H,B=H) images. The \textit{dashed boxes} in the \textit{upper panels} indicate the region enlarged in the \textit{lower panels}. Ellipses indicate where the flux is measured (on-source:\textit{solid}, off-source:\textit{dashed}). The ellipse pairs (on+off) is alphabetically tagged in order of increasing RA (cf.~\textit{lower-left} panel). Circles with a red slash indicate the region excluded during the flux measurement. The measured fluxes are listed in Table \ref{tbl-flux}.}
\end{figure}

\clearpage
\begin{figure}
\figurenum{5}
\centering
\includegraphics[scale=0.9]{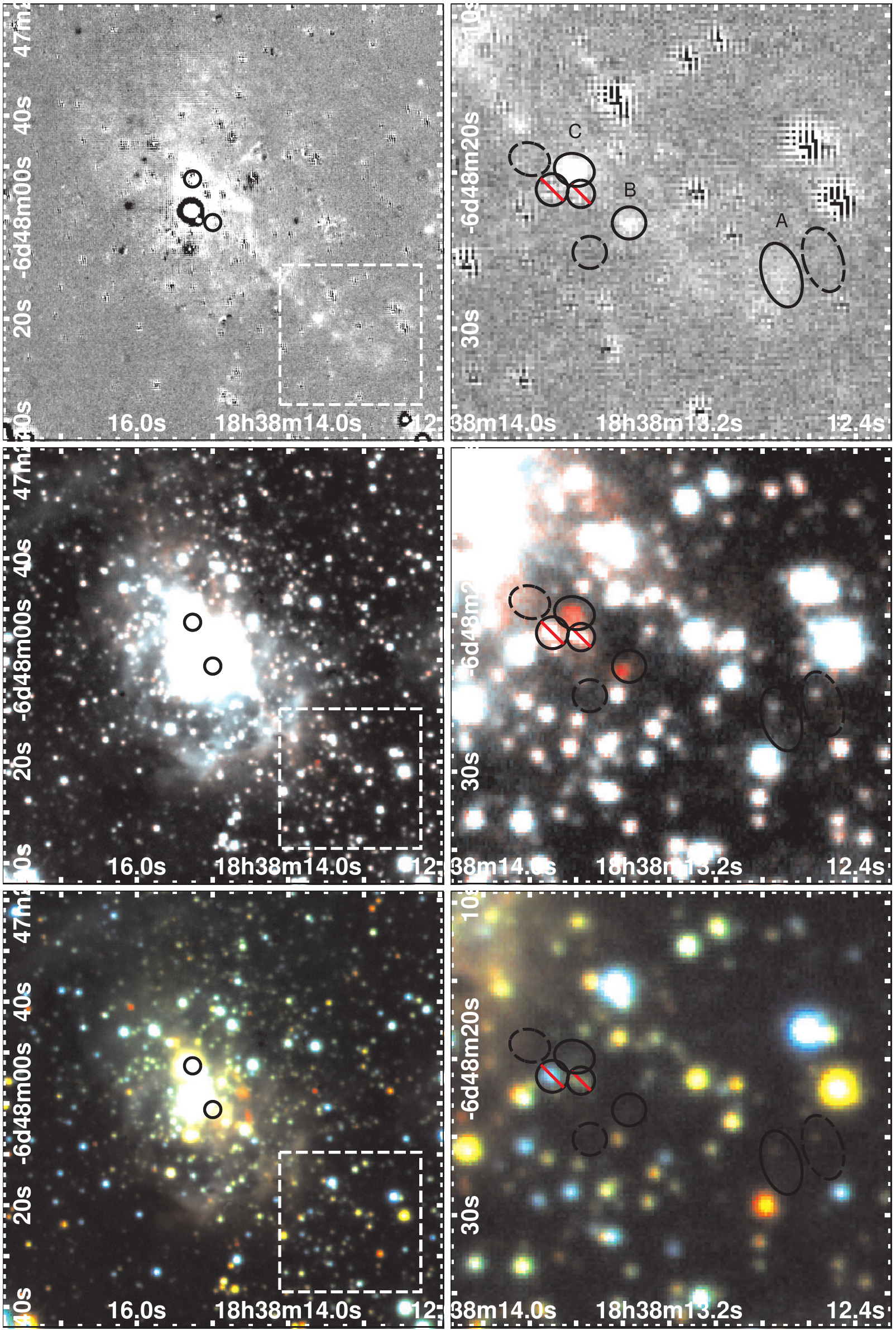}\\(a) G025.3824$-$00.1812 (\textit{upper-circle}) and G025.3809$-$00.1815 (\textit{lower-circle})
\caption{(Continued)}
%\caption{Outflow features around CORNISH sources seen in the continuum-subtracted [\astFe{II}] (\textit{left panels}) and RGB  composite (\textit{right panels};R=[\astFe{II}],G=H,B=H) images. The \textit{dashed boxes} in the \textit{upper panels} indicate the region enlarged in the \textit{lower panels}. Ellipses indicate where the flux is measured (on-source:\textit{solid}, off-source:\textit{dashed}). The ellipse pairs (on+off) is alphabetically tagged in order of increasing RA (cf.~\textit{lower-left} panel). Circles with a red slash indicate the region excluded during the flux measurement. The measured fluxes are listed in Table \ref{tbl-flux}.}
\end{figure}

\clearpage
\begin{figure}
\figurenum{5}
\centering
\includegraphics[scale=0.9]{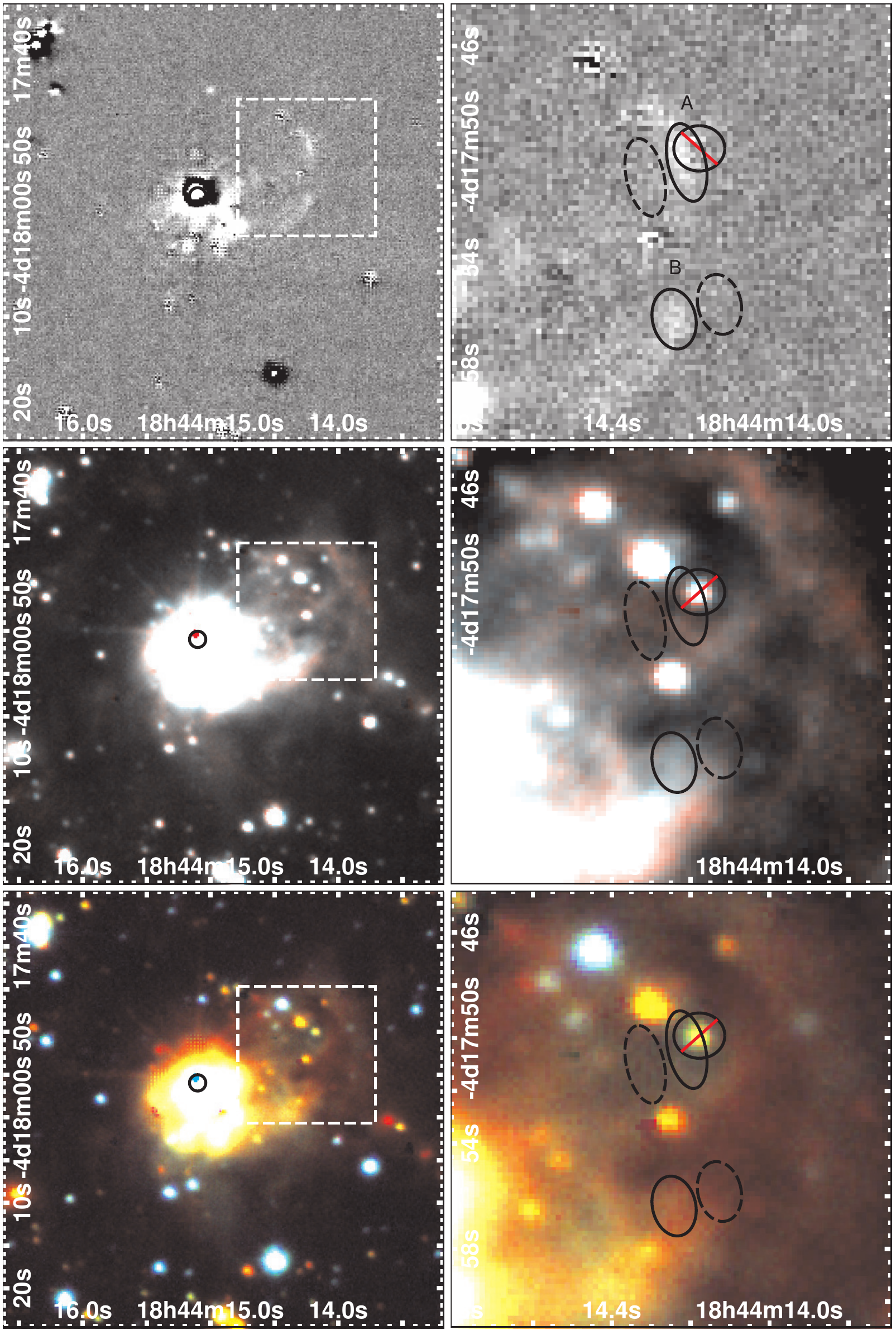}\\(b) G028.2879$-$00.3641
\caption{(Continued)}
\end{figure}

\clearpage
\begin{figure}
\figurenum{5}
\centering
\includegraphics[scale=0.9]{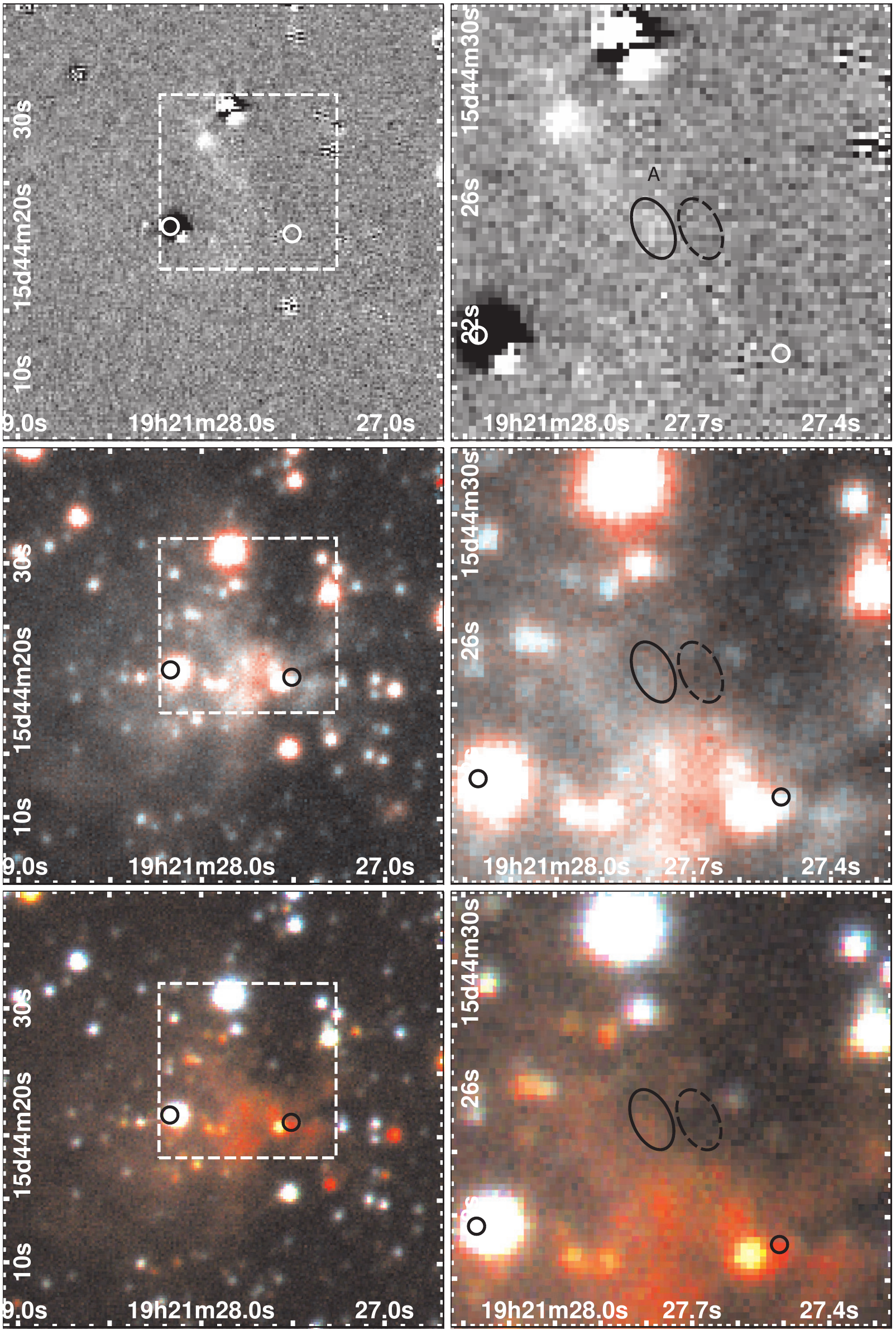}\\(c) G050.3157+00.6747 (\textit{left-circle}) and G050.3152+00.6762 (\textit{right-circle})
\caption{(Continued)}
\end{figure}

\clearpage
\begin{figure}
%\figurenum{3}
\centering
\includegraphics[scale=0.9]{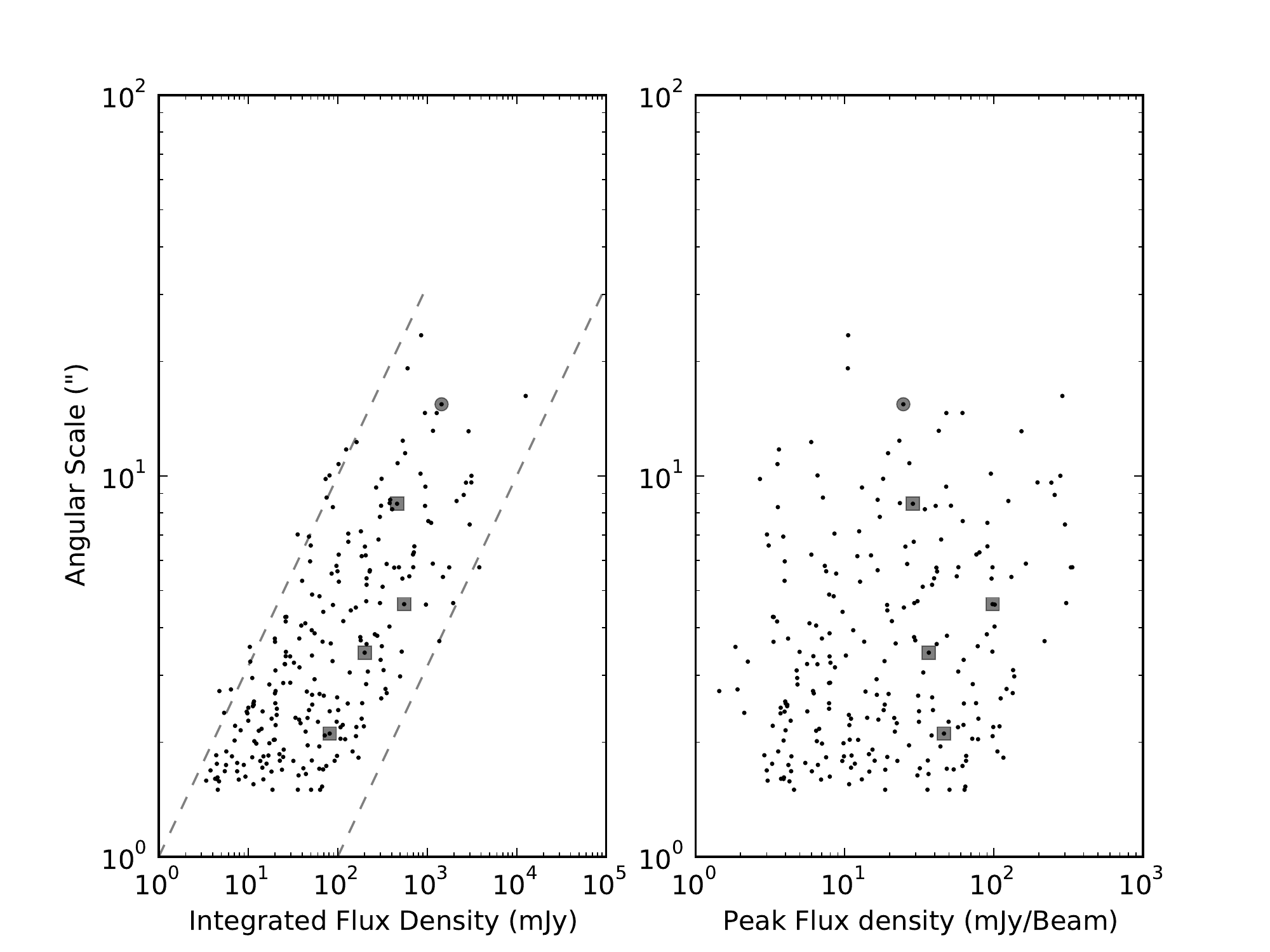}
\caption{
The UCHIIs scatter plot using the parameters from the CORNISH UCHII catalog. 
The \textit{points} are the UCHII candidates from the CORNISH catalog which provides the angular scale, integrated flux density, and peak flux density. 
%\mod{The \textit{squares} indicates the UCHIIs that shows the [\astFe{II}] outflow features.}
The \textit{squares} and \textit{circles} indicate the UCHIIs that show [\astFe{II}] outflow features and candidate features, respectively. 
%The \textit{red-filled} and \textit{black-filled} symbols indicate whether targets' [\astFe{II}] outflow features contact with the CORNISH 5 GHz features or not (\textit{red-filled}=no contact; \textit{black-filled}=in contact).
The \textit{gray dashed} lines in the \textit{left} plot show the locus of (angular scale) $\sim$ (integrated flux density)$^{1/2}$.
\label{fig-cor}
}
\end{figure}

%\clearpage
%\begin{figure}
%%\figurenum{3}
%\centering
%\includegraphics[scale=1.0]{f7.pdf}
%\caption{
%The relation between the outflow mass loss rate (Table \ref{tbl-out}) and the UCHII peak flux density (Table \ref{tbl-cor}).
%%The relation between the outflow mass loss rate (Table \ref{tbl-out-she} and \ref{tbl-out-str}) and the UCHII peak flux density (Table \ref{tbl-cor}).
%\label{fig-cor-mout}
%}
%\end{figure}

\clearpage
\begin{figure}
%\figurenum{3}
\centering
\includegraphics[scale=0.63]{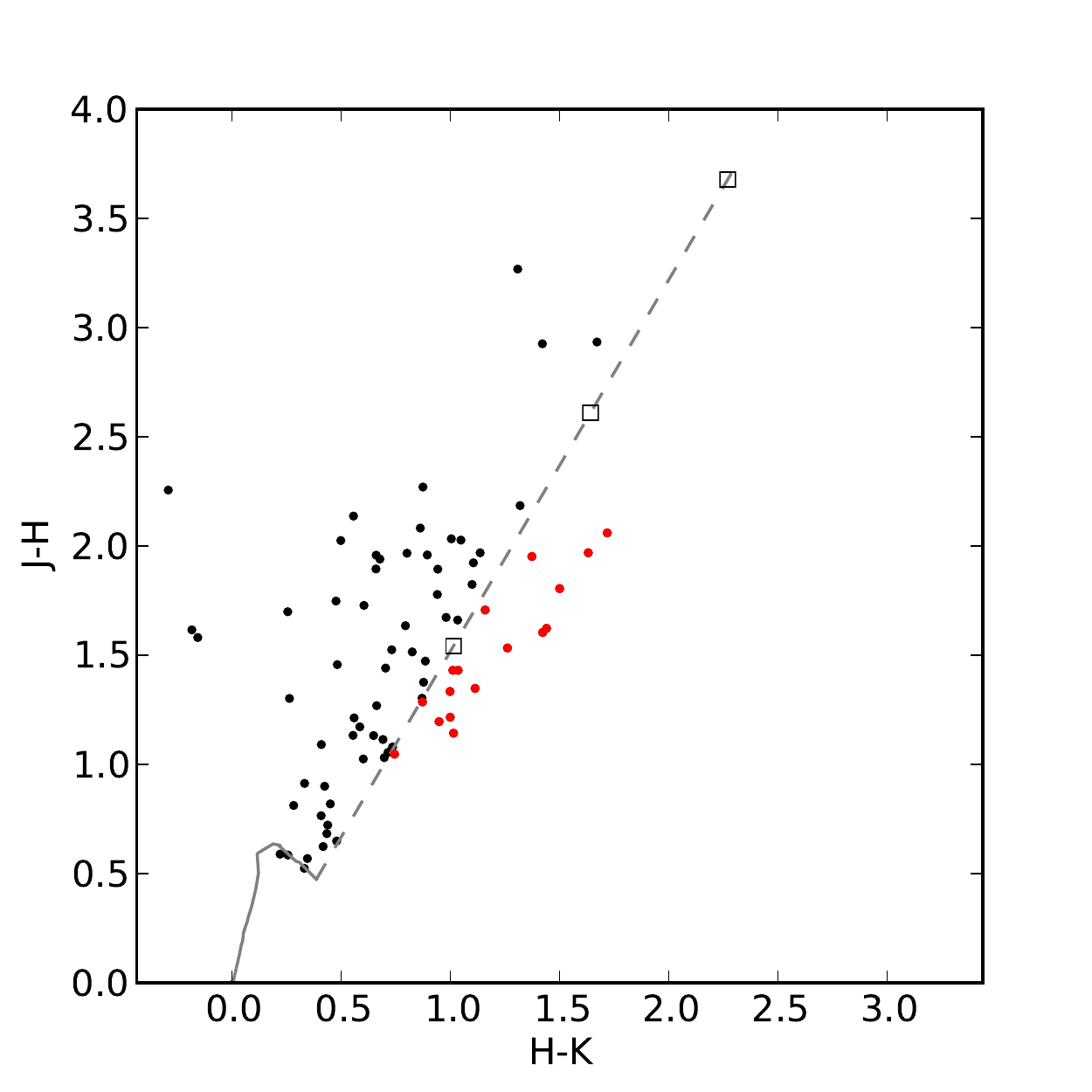}
\includegraphics[scale=0.45]{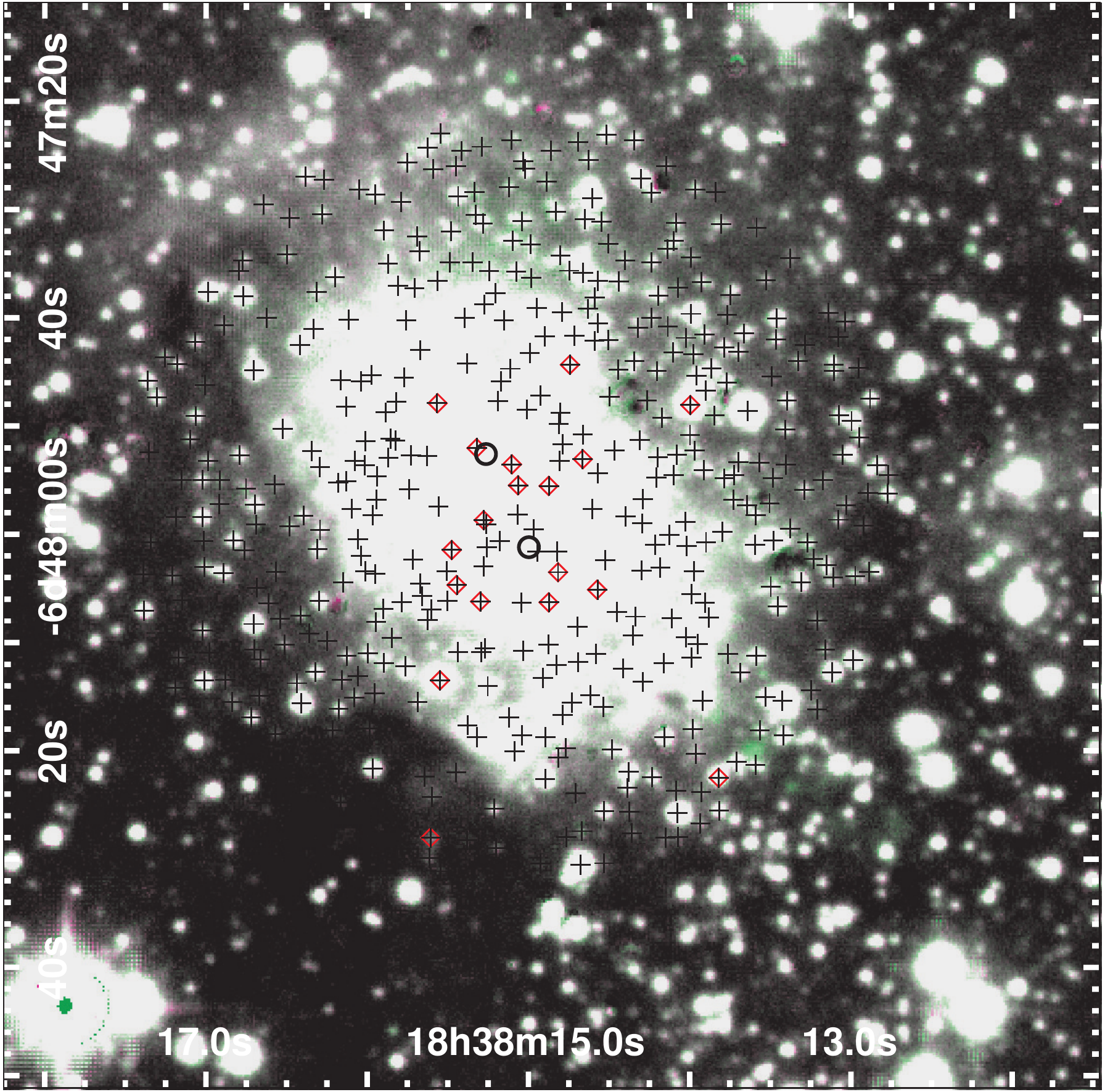}\\(a) G025.3824$-$00.1812 and G025.3809$-$00.1815
\caption{
\textit{left-panels}: color-color diagram.
The \textit{points} are the photometry data of the point sources seen in the inspected field.
Those points showing infrared excess are colored \textit{red}.
The photometry data in the panels (a) and (c) are from the UKIDSS GPS \citep{Lucas(2008)MNRAS_391_136}.
The photometry data in panel (b) are from 2MASS \citep{Skrutskie(2006)AJ_131_1163}, since no $J$-band photometry of UKIDSS GPS is available.
The \textit{solid-line} is the locus of main-sequence stars (B9-M6, \citealt{Hewett(2006)MNRAS_367_454}).
The \textit{dashed-line} is the reddening line \citep{Rieke(1985)ApJ_288_618}, and the \textit{squares} indicate the points that correspond to \Av=10, 20, and 30 from lower-left to upper-right.
\textit{right-panels}: RGB composite image of [\astFe{II}] ({green}) and H ({red}, blue) as in Figure \ref{fig-uchii}.
\textit{crosses} indicate the position of point sources plotted in the color-color diagram (\textit{left-panel}).
The point sources showing infrared excess are indicated with \textit{red-diamonds}.
{The contrast of the image for G025.3824$-$00.1812 and G025.3809$-$00.1815 is adjusted to show the [\astFe{II}] features, hence it looks different from Figure \ref{fig-uchii}a}.
\label{fig-ccd}
}
\end{figure}

\clearpage
\begin{figure}
\figurenum{7}
\centering
\includegraphics[scale=0.63]{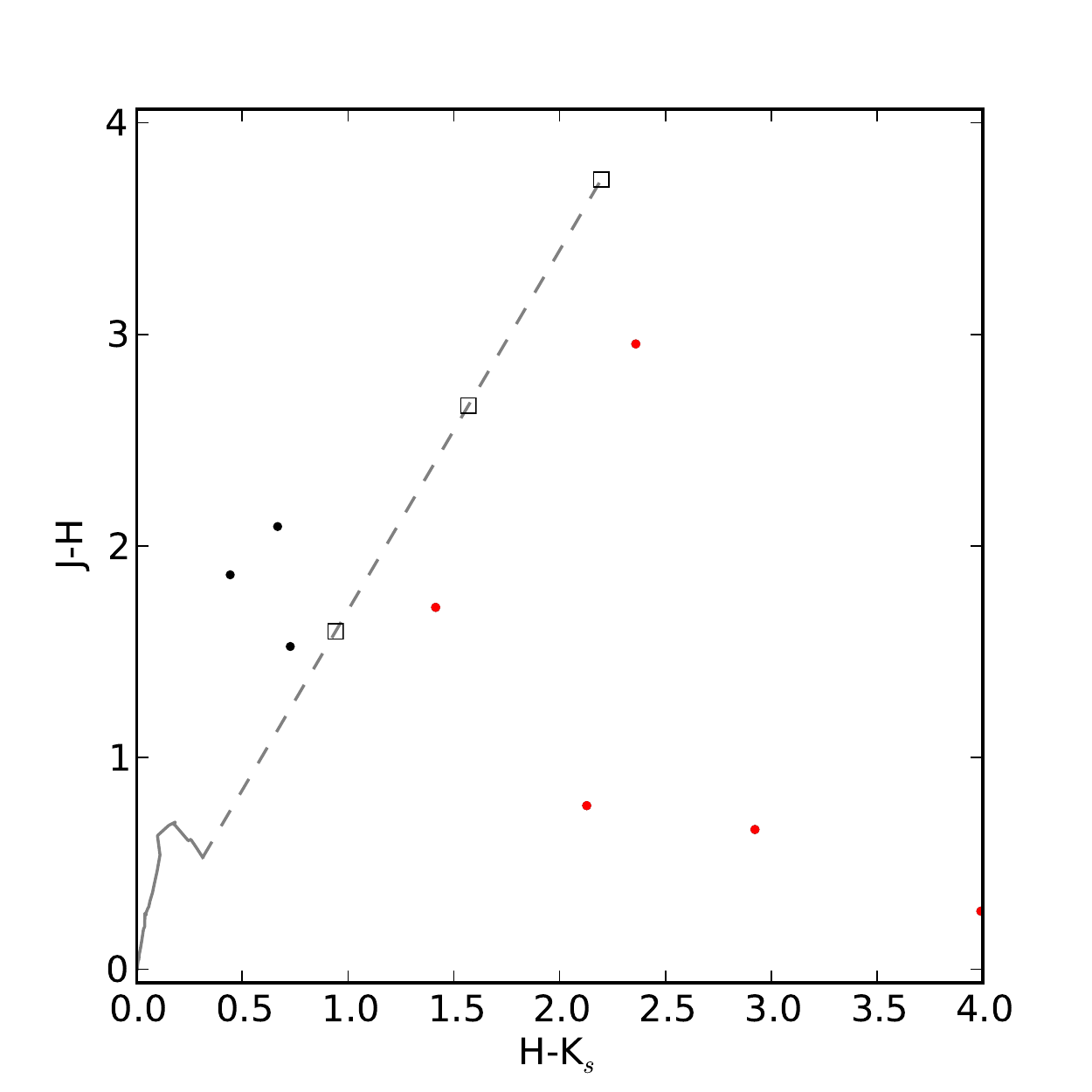}
\includegraphics[scale=0.45]{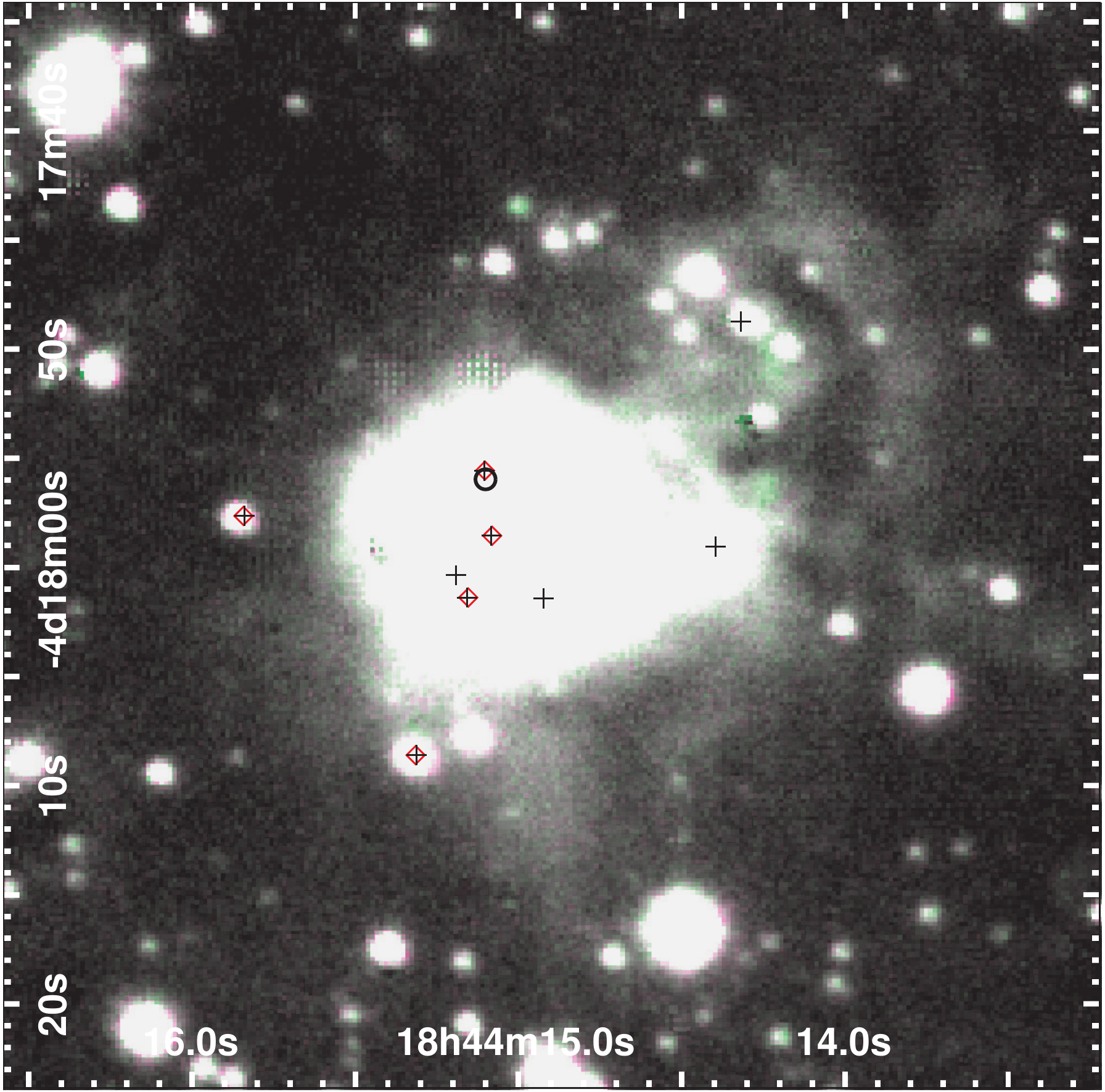}\\(b) G028.2879$-$00.3641
\caption{(Continued)}
\end{figure}

\clearpage
\begin{figure}
\figurenum{7}
\centering
\includegraphics[scale=0.63]{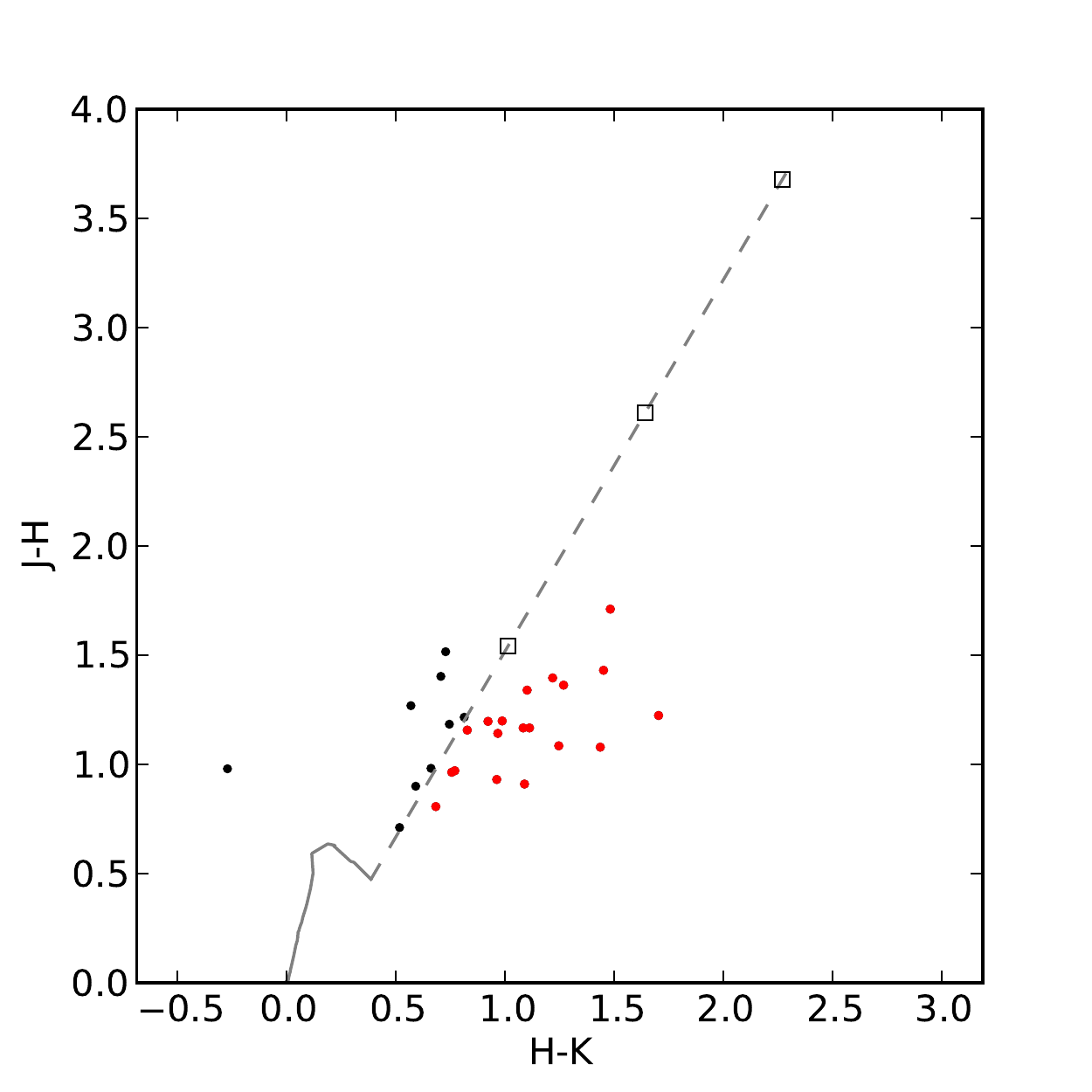}
\includegraphics[scale=0.45]{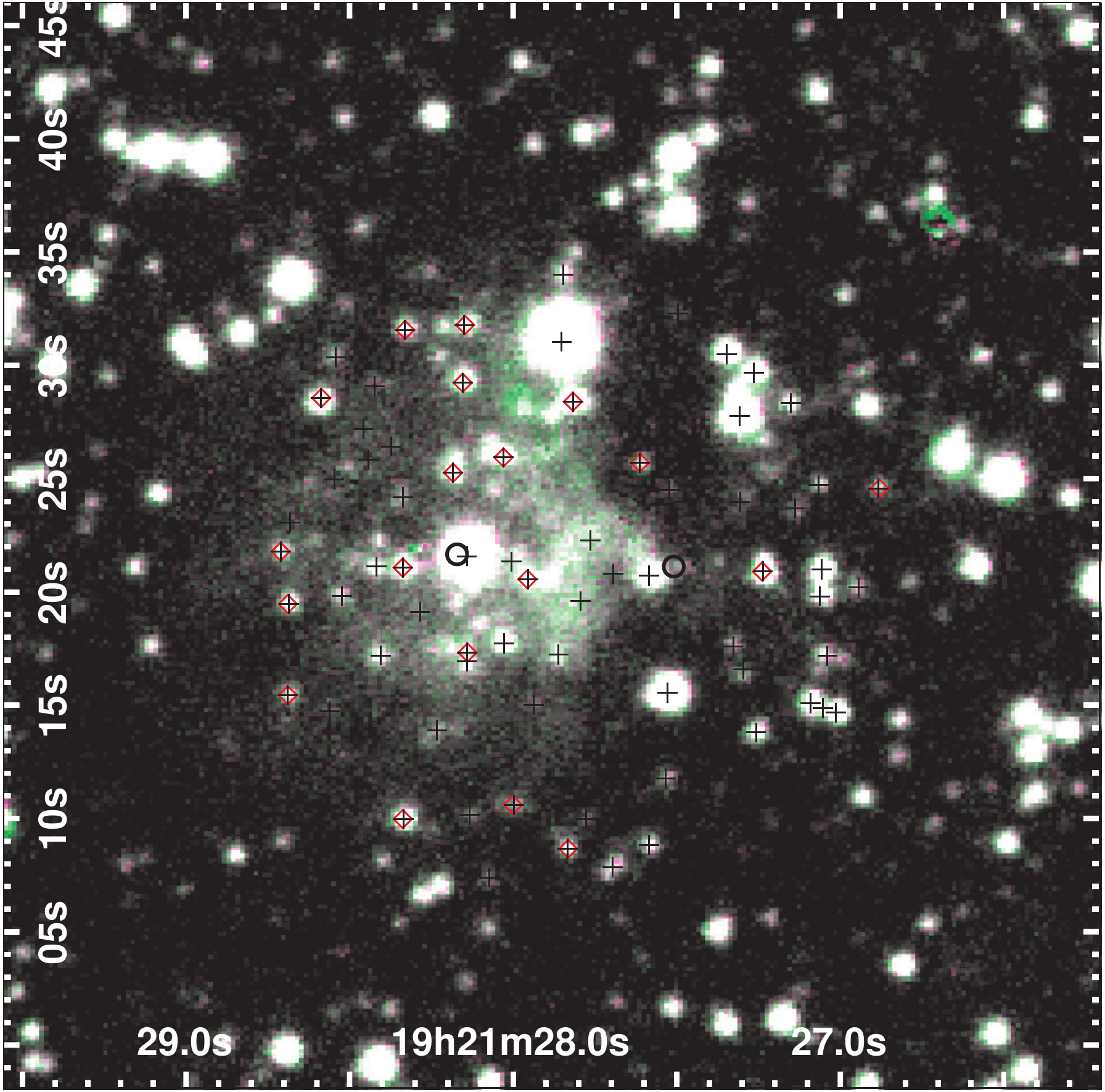}\\(c) G050.3157+00.6747 and G050.3152+00.6762
\caption{(Continued)}
\end{figure}

%%%%%%%%%%% TABLES
\clearpage
\begin{deluxetable}{ccrrr}
\tablewidth{0pt}
%\tabletypesize{\scriptsize}
%\tabletypesize{\small}
\tablecaption{Flux Measurements of [\astFe{II}] Features \label{tbl-flux}}
\tablehead{
\colhead{Region\tablenotemark{a}} & \colhead{RA, Dec (J2000)} & \colhead{\Av\tablenotemark{b}} & \multicolumn{2}{c}{[Fe II] flux} \\
& & & \colhead{Observed} & \colhead{Dereddened\tablenotemark{c}} \\
& & & \multicolumn{2}{c}{($10^{-15}$ erg s$^{-1}$ cm$^{-2}$)}
}
\startdata
G025.3809$-$00.1815-A & 18:38:12.719,$-$6:48:26.63&  8.9 &  5.2$\pm$0.2 & 23.9$\pm$0.9\\
G025.3809$-$00.1815-B & 18:38:13.373,$-$6:48:23.26&  8.9 &  2.8$\pm$0.2 & 12.6$\pm$0.7\\
G025.3809$-$00.1815-C & 18:38:13.608,$-$6:48:19.85&  8.9 &  6.6$\pm$0.2 & 30.0$\pm$0.9\\
G028.2879$-$00.3641-A & 18:44:14.210,$-$4:17:50.44&  8.7 &  2.5$\pm$0.1 & 10.9$\pm$0.5\\
G028.2879$-$00.3641-B & 18:44:14.242,$-$4:17:56.37&  8.7 &  6.1$\pm$0.2 & 26.6$\pm$0.8\\
G050.3152+00.6762-A & 19:21:27.788,+15:44:25.05& 20.1 &  2.2$\pm$0.1 & 65.8$\pm$2.4\\

\enddata
\tablenotetext{a}{For simplicity of the region name, we picked one UCHII for the name prefix when there were two UCHIIs relevant to the [\astFe{II}] features.}
\tablenotetext{b}{These values are inferred from near-infrared color excess. See section \ref{ana-res-flux} for detail.}
%\tablenotetext{b}{}
\tablenotetext{c}{The extinctions were corrected, using corresponding $A_V$ and the extinction curve of ``Milky Way, $R_V=3.0$'' \citep{Weingartner(2001)ApJ_548_296,Draine(2003)ARA&A_41_241}.}
%\tablenotetext{\dag3,\dag4}{These targets fall onto the overlapped regions (cf.~Fig.~\ref{fig-obs}). We measured their fluxes from each exposure, and averaged them.}
\end{deluxetable}

%\clearpage
%\begin{deluxetable}{cccr}
%\tablewidth{0pt}
%%\tabletypesize{\scriptsize}
%%\tabletypesize{\footnotesize}
%%\tabletypesize{\small}
%\tablecaption{Outflow Mass Loss Rate Estimated from the ``Fe-Shell'' Assumption \label{tbl-out-she}}
%\tablehead{
%\colhead{Targets} & \colhead{Velocity} & \colhead{Distance} & \colhead{\Mout} \\
%& \colhead{(\kms)} & \colhead{(kpc)} & \colhead{($10^{-5}$ $M_{\odot}$ yr$^{-1}$)}
%}
%\startdata
%\input{table2.tex}
%\enddata
%\tablenotetext{a}{We adopted the wind velocity of 100 \kms.}
%\tablenotetext{b}{The distance is adopted referring previous estimates. See text for detail.}
%\end{deluxetable}
%
%\clearpage
%\begin{deluxetable}{cccccr}
%\tablewidth{0pt}
%%\tabletypesize{\scriptsize}
%%\tabletypesize{\footnotesize}
%%\tabletypesize{\small}
%\tablecaption{Outflow Mass Loss Rate Estimated from the ``Fe-Stream'' Assumption   \label{tbl-out-str}}
%\tablehead{
%\colhead{Targets} & \colhead{Velocity} & \colhead{Distance} & \colhead{length} & \colhead{Solid Angle} & \colhead{\Mout} \\
%& \colhead{(\kms)} & \colhead{(kpc)} & \colhead{($''$)} & \colhead{($10^{-11}$ sr)} & \colhead{($10^{-6}$ $M_{\odot}$ yr$^{-1}$)}
%}
%\startdata
%\input{table3.tex}
%\enddata
%\tablenotetext{a}{We adopted the wind velocity of 100 \kms.}
%\tablenotetext{b}{The distance is adopted referring previous estimates. See text for detail.}
%\end{deluxetable}

\clearpage
\begin{deluxetable}{ccccccr}
\tablewidth{0pt}
%\tabletypesize{\scriptsize}
\tabletypesize{\footnotesize}
%\tabletypesize{\small}
\tablecaption{Outflow Mass Loss Rate Estimated from Dereddened [\astFe{II}] flux \label{tbl-out}}
\tablehead{
\colhead{Region\tablenotemark{a}} & \colhead{Velocity\tablenotemark{b}} & \colhead{Distance\tablenotemark{c}} & \colhead{length} & \colhead{Solid Angle} & \colhead{Method\tablenotemark{d}} & \colhead{\Mout\tablenotemark{e}} \\
& \colhead{(\kms)} & \colhead{(kpc)} & \colhead{($''$)} & \colhead{($10^{-11}$ sr)} &  & \colhead{($10^{-6}$ $M_{\odot}$ yr$^{-1}$)}
}
\startdata
G025.3809$-$00.1815-A & 200& 3.9 & \nodata & \nodata & Sh & 4.44$\pm$0.17\\
G025.3809$-$00.1815-B & 100& 3.9 & \nodata & \nodata & Sh & 16.40$\pm$0.90\\
G025.3809$-$00.1815-C & 100& 3.9 & \nodata & \nodata & Sh & 39.00$\pm$1.13\\
G028.2879$-$00.3641-A & 200& 3.2 & \nodata & \nodata & Sh & 1.36$\pm$0.07\\
G028.2879$-$00.3641-B & 200& 3.2 & \nodata & \nodata & Sh & 3.32$\pm$0.10\\
G050.3152+00.6762-A & 200& 9.2 & 1.0 & 4.3 & St & 8.87$\pm$0.33\\

\enddata
\tablenotetext{a}{For simplicity of the region name, we picked one UCHII for the name prefix when there were two UCHIIs relevant to the [\astFe{II}] features.}
\tablenotetext{b}{We adopted the outflow velocity of 100 \kms{} {or 200 \kms{}. See Section \ref{ana-res-mout} for detail.}}
\tablenotetext{c}{The distance is adopted from previous estimates. See Section \ref{ana-res-mout} for detail.}
\tablenotetext{d}{``Sh'' and ``St'' indicate the ``Fe-Shell'' and ``Fe-Stream'' methods used for the \Mout{} estimation, respectively. See Section \ref{ana-res-mout} for detail.}
\tablenotetext{e}{\mod{These values are subject to substantial uncertainties originating from the uncertain outflow velocity, extinction, Fe depletion, etc. See the text for details. The listed errors are from the [\astFe{II}] flux error only.}}
\end{deluxetable}

\clearpage
\begin{deluxetable}{crrrr}
\tablewidth{0pt}
%\tabletypesize{\scriptsize}
\tabletypesize{\footnotesize}
%\tabletypesize{\small}
\tablecaption{UCHII Parameters from the CORNISH Catalog \label{tbl-cor}}
\tablehead{
\colhead{UCHII} & \colhead{Peak Flux Density} & \colhead{Integrated Flux Density} & \colhead{Angular Scale} & \colhead{Deconvolved Size}\\
& \colhead{(mJy/beam)} & \colhead{(mJy)} & \colhead{($''$)} & \colhead{($''$)}
}
\startdata
G013.8726+00.2818\tablenotemark{a} &    24.71$\pm$2.21 &  1447.55$\pm$129.84 &   15.430$\pm$0.009 &     15.4\\
G025.3809$-$00.1815 &    28.64$\pm$2.57 &   460.83$\pm$42.66 &    8.456$\pm$0.013 &      8.3\\
G025.3824$-$00.1812 &    36.62$\pm$3.32 &   200.13$\pm$20.03 &    3.436$\pm$0.013 &      3.1\\
G028.2879$-$00.3641 &    98.00$\pm$8.73 &   552.77$\pm$51.90 &    4.607$\pm$0.005 &      4.4\\
G050.3152+00.6762 &    46.28$\pm$4.12 &    81.31$\pm$8.07 &    2.107$\pm$0.007 &      1.5\\

\enddata
\tablenotetext{a}{This UCHII might have [\astFe{II}] features around it, but with a low plausibility (cf.~Section \ref{ana-res-det} and Figure \ref{fig-uchii-cand}).}
%\tablenotetext{b}{The distance is adopted referring previous estimates. See text for detail.}
\end{deluxetable}

\clearpage
\begin{deluxetable}{ccc}
\tablewidth{0pt}
%\tabletypesize{\scriptsize}
%\tabletypesize{\small}
\tablecaption{Dynamical Time Comparison\tablenotemark{a} \label{tbl-time}}
\tablehead{
\colhead{UCHII} & \colhead{\astH{II} Region} & \colhead{Outflow	}\\
\colhead{} & \colhead{Expansion Time} & \colhead{Travel Time}\\
& \colhead{($10^3$ yr)} & \colhead{($10^3$ yr)}
}
\startdata
G025.3809$-$00.1815 & 5.1 & 4.0$-$6.1\tablenotemark{b}\\
G025.3824$-$00.1812 & 1.9 & 4.7$-$7.6\tablenotemark{b}\\
G028.2879$-$00.3641 & 2.2 & 1.0$-$1.1\tablenotemark{b}\\
G050.3152+00.6762 & 2.2 & 1.3\\

\enddata
\tablenotetext{a}{These quantities are estimated with several assumptions. See Section \ref{dis-nat} for detail.}
\tablenotetext{b}{The time is estimated for all the relevant [\astFe{II}] features (cf.~Figure \ref{fig-uchii}).}
%\tablenotetext{c}{The time is estimated for the closest [\astFe{II}] feature, B (cf.~Figure \ref{fig-uchii}b).}
\end{deluxetable}

\end{document}